\documentclass{ws-ijqi}
\usepackage{amssymb,amscd,graphicx}
\usepackage{subfigure}
\usepackage{color}

\usepackage{graphics}
\usepackage{bm}
\usepackage{bbm} 
\usepackage{amsfonts}
\usepackage{mathrsfs}
\usepackage{multirow}

\usepackage{makeidx}\makeindex

\newcommand{\kb}[2]{\ensuremath{\vert #1 \rangle \langle #2 \vert}}
\def\bra#1{\mathinner{\langle{#1}|}}
\def\ket#1{\mathinner{|{#1}\rangle}}

\newcommand{\id}{\mathbbm{1}}

\def\id{{\mathchoice {\rm 1\mskip-4mu l} {\rm 1\mskip-4mu l} {\rm 1\mskip-4.5mu l} {\rm 1\mskip-5mu l}}}

\begin{document}

\title{AN INTRODUCTION TO ONE-WAY QUANTUM COMPUTING\\ IN DISTRIBUTED ARCHITECTURES}

\author{EARL T. CAMPBELL}\address{Department of Physics and Astronomy, University College London, WC1E 6BT, U. K.}
\author{JOSEPH FITZSIMONS}\address{Department of Materials, University of Oxford, OX1 3PH, U. K. \\ Institute for Quantum Computing, University of Waterloo, Waterloo, Ontario, Canada}

\maketitle

\begin{abstract}
This review provides a gentle introduction to one-way quantum computing in distributed architectures. One-way quantum computation shows significant promise as a computational model for distributed systems, particularly those architectures which rely on probabilistic entangling operations. We review the theoretical underpinnings of one-way quantum computation and discuss the practical issues related to exploiting the one-way model in distributed architectures. 
\end{abstract}

\section{Introduction}

The seminal work of Raussendorf and Briegel\cite{RB01a} proposed the fundamentals of \textit{one-way quantum computing}, and these ideas were developed by follow-up papers from numerous sources\cite{RBB01a,HEB01a,N02a}. One-way quantum computing is so-called because it is driven by irreversible measurements performed on a large entangled state, rather than by reversible gates.  Although the gate-based model of quantum computing uses entangling operations, these are spread throughout the computation, whereas in one-way quantum computing all the entanglement required is present in the initial state\cite{VandenNest06,NDMB01a}.
  
Many recent architectures proposed for distributed quantum computing suffer from an inability to deterministically entangle qubits. If entangling gates between single qubit nodes are projective and probabilistic then a naive application of the gate-based approach will be prohibitively inefficient.  However, projective and probabilistic entangling operations can be employed quite naturally to \textit{grow} an entanglement resource required for the one-way model\cite{N01a,DR01a,KRE01a,ACL01a,BEFMMS01a}.  Conveying an understanding of how one grows an entanglement resource is our core goal here.

The entanglement resources of one-way quantum computing are called \textit{graph}, or \textit{cluster}, states and our starting point is to explain what these entanglement resources are.   We introduce two invaluable tools for describing graph states.  Firstly, we outline the stabilizer formalism, which uses operators to describe quantum states.  Secondly, we introduce a graphical representation from which graph states inherited their name.  Having described graph states, their information processing capabilities will be introduced by analogy with teleportation. Graph states will be shown to be capable of universal quantum computing in a deterministic fashion. The determinism of one-way quantum computing will be a point of interest as it is this feature that was not obvious prior to the work of Raussendorf and Briegel. Indeed, since the measurements on graph states necessary to perform computation have random outcomes it is counter-intuitive that this evolution can be harnessed to deterministically process information. As we shall see, certain measurements on graph states can be classically simulated, and so can be removed to yield a smaller, more efficient resource state for a given algorithm.

Having established the basics of one-way quantum computing, we move on to describe how it can be physically embodied in a \textit{distributed} quantum computer.  An architecture for quantum computing is distributed whenever it has a natural partitioning into nodes, small quantum registers, that are interconnected by some network that allows communication between nodes. Each node is a quantum computer of limited size, where quantum control is excellent between that small number of qubits. This is an apt description of many next generation proposals\cite{ODH01a,JTSL02a}, where an initial design has been found to work well on a small scale, but scaling up the size of a node is either difficult or simply impossible. Rather than abandoning an excellent small scale technology, the distributed paradigm prescribes that we chain up these technologies into a network of nodes. The approach is proving attractive to experimental groups because separation of nodes allows both for more efficient cooling and individual addressability, which become possible by virtue of spatial separation.  In solid states systems with densely packed qubits addressing individual qubits remain a significant obstacle making the distributed paradigm an appealing alternative.

The concept of a distributed quantum computer (DQC) is quite general, but it is especially suited to so called \textit{hybrid} matter-optical systems.  In hybrid systems some collection of matter qubits form a node, and the inter-node entanglement is generated by an optical mechanism\cite{CCFZ01a,BKPV01a,FZLGX01a,BPH01a,BK01a,LBK01a,B01a,BES01a,BBFM01a,LBBKK01a,CB01a}.  Such schemes have been experimentally demonstrated between: ions held in separate electromagnetic traps\cite{Beugnon06,MMOYMDM1a,Olmschenk09}; and clouds of cold caesium atoms in separate containers\cite{CRFPEK01a}.  We will see that when photon detectors register a particular set of detector ``clicks", this can herald the creation of entanglement if the photon source is unknown.  Typically, information regarding a photon's path to the detector is erased by use of a beam splitter, so that it cannot be determined which node was the emitter.   Our introduction to hybrid systems is split into a two parts: firstly, a review of the underlying physics of atom-light interactions and linear optical transformations; secondly we outline a number of techniques for dealing with practical issues which arise in such systems.  Since photons are easily lost, for example by absorption or detector inefficiency, robustness against this problem will play a prominent role in our review of these protocols.

One-way quantum computing and distributed quantum computing are partially independent topics.  The former is a model of computation, and the latter a physical design.  However, they fit together neatly to offer a complete approach to quantum computing.  On one hand DQC gives us a method of producing entanglement based on photon measurements that are inherently probabilistic and projective.  On the other hand, one-way quantum computing requires an entanglement resource that is simple to describe.   Fortunately,  several strategies exist that ensure these imperfect entangling operations can efficiently grow the desired graph states, and we outline these in section~\ref{sec:Growth_strategies}.  We close with a brief discussion of how to tolerate other imperfections using fault tolerance techniques and entanglement distillation.

\section{One-way computation}

\subsection{The stabilizer formalism}

The stabilizer formalism\cite{Got96a} is a powerful tool for describing certain quantum states, and achieves this by a compact description in terms of operators instead of wavefunctions.  The stabilizer formalism was originally developed in the context of quantum error correction, with most of the original literature emphasizing this application. Our presentation of the stabilizer formalism is stripped of these details, providing just the essentials for understanding one-way quantum computing.

For an operator description of a quantum state, $\vert \psi \rangle$, it is sufficient to have: (\textit{i}) a \textit{non-degenerate} operator, $Q$, for which $\vert \psi \rangle$ is a eigenfunction; and (\textit{ii}) a number, $q_{i}$, that specifies the correct eigenvalue, $Q\vert \psi \rangle = q_{i}\vert \psi \rangle$.  Non-degeneracy is the property of an operator that has only one eigenfunction for each eigenvalue.  Hence, a single such operator is sufficient to identify a unique state.  For simplicity, the eigenvalue can be absorbed into the operator to make a new operator $S=(1/q_{i})Q$.  For now, we will call $S$ a \textit{stabilizing} operator of $\ket{\psi}$, similarly for any other operator satisfying $S\ket{\psi}=\ket{\psi}$.   Later we further restrict the class of stabilizing operators.

Unfortunately, as with the wavefunction, this non-degenerate stabilizing operator may not have a simple decomposition.  Instead, consider what can be deduced about $\vert \psi \rangle$ from knowing that it is stabilized by a degenerate operator $S$, with $j$ orthogonal eigenfunctions $\left\{ \vert \psi_{1} \rangle, \vert \psi_{2} \rangle,... \vert \psi_{j} \rangle   \right\}$ with eigenvalue $+1$.  It is clear that $\vert \psi \rangle$ may be any of these eigenfunctions, or indeed any complex linear combination of them; that is, $\vert \psi \rangle$ is within the $j$-dimensional subspace, $\mathscr{H}_{S}$, stabilized by $S$.  Given a set of stabilizing operators $\mathcal{M}=\left\{ S_{1}, S_{2},... S_{n} \right\}$, $\vert \psi \rangle$ must be within the subspace stabilized by all these operators.  Therefore, given enough stabilizing operators, even if each is degenerate, we can reduce the number of possible states to one definite state. This set of stabilizing operators, $\mathcal{M}$, forms the stabilizer of $\ket{\psi}$. Typically, we consider only stabilizing operators chosen from tensor products of Pauli operators ($X$, $Y$ and $Z$), the Pauli group.  By restricting stabilizing operators to Pauli operators we limit ourselves to being able to describe a subset of all possible states, which we call \textit{stabilizer states}.  In exchange for this limitation, the stabilizer formalism gains a conciseness that enables it to avoid the exponential number of variables required to describe a general quantum state.

\subsubsection{Commutation of stabilizing operators}Any two stabilizing operators of a state must commute on the state-space they stabilize, since

\begin{eqnarray}
[S_j(\psi),S_k(\psi)] \ket{\psi} & = & S_j(\psi) S_k(\psi)\ket{\psi} - S_k(\psi) S_j(\psi)\ket{\psi} \\
& = & S_j(\psi) \ket{\psi} - S_k(\psi) \ket{\psi} \\
& = & \ket{\psi} - \ket{\psi} = 0.
\end{eqnarray}

Further, for stabilizing operations in the Pauli group this implies that the operators themselves must commute.

\subsubsection{Products of stabilizing operators}

For any two stabilizing operators, $S_k$ and $S_\ell$, their product must also be in the stabilizer of the same state, since:
\begin{eqnarray}
(S_k S_\ell) \ket{\psi} &=& S_k (S_\ell \ket{\psi}) \\
&=& S_k \ket{\psi} \\
&=& \ket{\psi}.
\end{eqnarray}
As a consequence of this, in order to track the evolution of a quantum state, it is not necessary to individually track the evolution of all operators in the stabilizer. Instead it is possible to track only a subset of stabilizing operators, which by multiplication generates the rest of the stabilizer. Any minimal collection of operators that suffices to describe the stabilizer are called \textit{generators} of the stabilizer.  It has been shown\cite{NC01b} that $n$ generators are sufficient to describe all $n$-qubit stabilizer states.

\subsubsection{Unitary evolution\label{stab-evo}}  Closed quantum systems, which are not interacting with an environment, are governed by the Schr\"{o}dinger equation.   After at time $\Delta$ of evolving under the Schr\"{o}dinger equation, an initial state $\ket{\psi(t)}$ evolves unitarily to $\ket{\psi(t+\Delta)} = U \ket{\psi(t)}$ where $U$ is a unitary operator satisfying $U^{\dagger}U=\id$.   If the initial state was stabilized by $S$, then:
\begin{eqnarray}
\ket{\psi(t+\Delta)} &=& U \ket{\psi(t)}\\
&=& U S \ket{\psi(t)} \\
&=& (U S U^\dagger) U \ket{\psi(t)} \\
&=& (U S U^\dagger) \ket{\psi(t+\Delta)}
\end{eqnarray}
and so the new state $\ket{\psi(t+\Delta)} = U \ket{\psi(t)}$ is stabilized by $S' = U S U^\dagger$.  In general, $U$ will not map a Pauli operator to a Pauli operator, and may evolve an initial stabilizer state into some state which does not admit an efficient stabilizer description.  However, some unitaries will always map Pauli operators to Pauli operators.  Unitary operators that satisfy this property are said to belong to the Clifford group, $\mathcal{C}$, where the group structure entails that if $C_{1} \in \mathcal{C}$ and $C_{2} \in \mathcal{C}$  then the product is also in the Clifford group $C_{1}C_{2} \in \mathcal{C}$. Consequently, we can describe the whole Clifford group by describing a subset of operators that generate the rest of the group.  One collection of generators are the single qubit Hadamard, square-root phase flip and two-qubit control-phase gate:
\begin{eqnarray}
\label{EQN:Clifford_gates}
	H_{j} & = & \kb{0}{+}_{j}+\kb{1}{-}_{j} , \\ \nonumber
	P_{j} & = & \kb{0}{0}_{j}+i\kb{1}{1}_{j} ,\\ \nonumber
	CZ_{i}^{j} & = & \kb{0,0}{0,0}_{i,j}+ \kb{1,0}{1,0}_{i,j}+ \kb{0,1}{0,1}_{i,j}+ -\kb{1,1}{1,1}_{i,j} ,
\end{eqnarray}

where in the definition of the Hadamard, $H$, we have used the shorthand $\ket{\pm}=(\ket{0}\pm\ket{1})\sqrt{2}$. The Clifford group is an important subgroup of unitary operations, containing gates necessary to demonstrate many of the important features of quantum mechanics, including non-commuting operators, manipulation of quantum superpositions and creation of entanglement. One of the most striking features of this group of operations, however, is that unlike general unitary operations evolution under Clifford group operations often remains classically simulable, a result known as  the Gottesman-Knill theorem\cite{G02a,G01a}.

\subsubsection{Measurements} 

What measurements can we encompass within the stabilizer formalism?  If we measure an observable, $M$, and get outcome $m_{i}$, then the measured system is projected into a state stabilized by $m_{i}^{-1}M$.  Hence, we expect that measurements of Pauli operators can be modeled within the stabilizer formalism.  Clearly,  $m_{i}^{-1}M$ is added to the stabilizer of the state, but does the projection remove any operators from the stabilizer?  Recall that all operators within the stabilizer must commute with each other, and that all Pauli operators either commute or anti-commute.  It follows that projection will cause all operators that anti-commute with $m_{i}^{-1}M$ to be eliminated from the stabilizer.  In terms of stabilizer generators, there is always a choice of generators where no more than one generator anti-commutes with the measurement operator.

In addition to simulating Pauli measurement projections we can also predict the outcome probabilities.  The are two different cases: when the measurement operator commutes with the whole stabilizer group, and when it does not.  If it commutes with the stabilizer the system must already be in an eigenstate of the observable, and one outcome occurs with unit probability.  When this is not the case, the outcomes $+1$ and $-1$ occur with equal probability.

\subsubsection{The Gottesman-Knill theorem\label{sec:GKT}}

We have loosely remarked that stabilizer formalism provides a compact description of stabilizer states.  More formally, an n-qubit stabilizer state can be described by a classical bit string that scales as a polynomial in $n$.  We can arrive at this result by a simple counting argument:  An $n$-qubit stabilizer state is described by $n$ generators, so we require $n{G}$ bits where $G$ is the number of bits per generator.  Each generator consists of a $\pm$ sign (1 bit) and a tensor product of $n$ Pauli operators ($2n$ bits\footnote{The factor of 2 arises because we use 2 bits to describe each Pauli operator, e.g. $\id=(0,0)$, $X=(0,1)$, $Y=(1,1)$, $Z=(1,0)$.}) , so $G=2n+1$.  The grand total is $n(2n+1)$ bits, which is a slow scaling polynomial much smaller than the exponential amount of information required to describe a general quantum state.  This argument suffices to establish the efficiency of the stabilizer formalism, though further reductions have been investigated\cite{aaronson-2004,AB01a}. 

This result lies at the heart of the Gottesman-Knill theorem. The theorem states that any quantum circuit composed entirely of gates in the Clifford group and Pauli basis measurements, taking as input a stabilizer state, can be simulated on a classical computer with only polynomial overhead. Formally the theorem can be stated as follows:

\begin{theorem}
 Any quantum computer performing only: a) initialisation of qubits in stabilizer states, b) Clifford group gates, c) measurements of Pauli group operators, and d) Clifford group operations conditioned on classical bits, which may be the results of earlier measurements, can be perfectly simulated in polynomial time on a probabilistic classical computer. \end{theorem}

On the basis of the preceding sections, this should seem plausible to the reader. One way to see that the Gottesman-Knill theorem must hold is to consider the effect of such a circuit on the stabilizer of the initial quantum state. This initial state has by definition an efficient stabilizer description. In general, measurements can be postponed until after the unitary part of a quantum circuit, by adding a single ancillary qubit per measurement.

From section \ref{stab-evo} the stabilizing operators, $\{S_k\}$, of the initial quantum state evolve under an operator $U$ as $U S_k U^\dagger$. If $U$ is an element of the Clifford group and $\{S_k\}$ are Pauli operators, then the resultant generators will also be within the Pauli group. Thus the quantum state produced as a result of the circuit must also be a stabilizer state. As the size and form of the set of generators of the stabilizer describing the quantum state remains constant, the number of parameters describing the state also remain constant. This is in stark contrast to the general case, where the number of free parameters describing a quantum state can grow exponentially with the number of quantum gates. For a formal proof of the theorem and exact computational resources required for such a simulation we refer to the relevant literature\cite{G02a,G01a,aaronson-2004,AB01a}.

\subsection{Graph states}
\subsubsection{The constructive definition}

In mathematics, a \textit{graph} is a particular kind of diagram composed of points (\textit{graph vertices}) connected by lines (\textit{graph edges}), where some simple examples are shown in tables~{\ref{ExamGraphs}} and~{\ref{EqGraphs}}.  In quantum information graph states are a particular class of quantum states that can be represented by a graph.  There are numerous definitions of a graph state, but the simplest definition to grasp is the constructive definition, where a graph specifies a procedure for constructing the corresponding quantum state.  Firstly, each graph vertex, \textit{i}, represents a qubit prepared in the state $\ket{+}_{i}$.  Secondly, each graph edge between two vertices, \textit{i} and \textit{j}, represents the application of a two-qubit controlled-phase gate, $CZ_{i}^{j}$ (defined in Eqn.~(\ref{EQN:Clifford_gates})).  All $CZ$ operations commute, and $CZ$ is unchanged on interchange of the control and target qubit.  Therefore, the edges do not require a time ordering or bias toward one qubit.  

Cluster states are a specific class of graph state for which the corresponding graph is a regular square lattice. These were initially introduced by Briegel and Raussendorf in the context of studying many-body entanglement\cite{BR01a}, but as we shall see in section \ref{clustergates}, also provide a universal resource for measurement based computation.  Although the constructive definition specifies a unique quantum state for each graph, the corresponding wavefunction rapidly becomes cumbersome as the graph grows in size.

\begin{table}[t]

\tbl{Three representations of graph states for: (\textit{a}) two-qubit linear chain; (\textit{b}) 4-qubit linear chain; (\textit{c}) a 5-qubit non-trivial graph.\label{ExamGraphs}}{  \normalsize
\begin{tabular}{|l | r c l | l |} \hline
 & & & & \\ 
Graph & & & Wavefunction & Stabilizer\\
representation &  & & representation & generators \\ 
\small
& & & & \\ \hline 
(a)& & & & \\
\multirow{2}{*}{\includegraphics{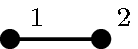} } 
& $\ket{\mathscr{G}_{a}}$ & $=$ & $CZ^{1}_{2}\ket{++}$ & $X_{1}Z_{2},$\\
  																		&  												& $=$ & $(1/\sqrt{2})(\ket{0+}+\ket{1-})$ & $ X_{2}Z_{1}$ \\ 
  																		& & & & \\ \hline 
 (b) 																		& & & & \\
\multirow{4}{*}{\includegraphics{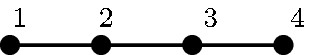} }   																		
  																		 & $\ket{\mathscr{G}_{b}}$ & $=$ & $CZ^{1}_{2}CZ^{2}_{3}CZ^{3}_{4}\ket{++++}$ & $X_{1}Z_{2},$ \\
 																			&	 & $=$ & $(1/\sqrt{2}) \big( \ket{+00+} + \ket{+01-}$  & $ X_{2}Z_{1}Z_{3},$  \\ 
 																																		&	 &  & $\ket{-10+} + \ket{-11-} \big)$ &  $X_{3}Z_{2}Z_{4},$ \\
 																																		&	 &  &  & $X_{4}Z_{3}$ \\ 
 																																		& & & & \\ \hline 
 		(c)																																& & & & \\
\multirow{4}{*}{\includegraphics{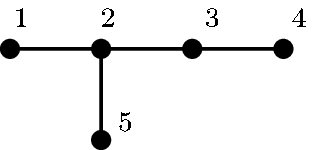} } 																																		 & $\ket{\mathscr{G}_{c}}$&$=$& $CZ^{1}_{2}CZ^{2}_{3}CZ^{2}_{5}CZ^{3}_{4}\ket{+++++} $  & $X_{1}Z_{2},$ \\ 
& &$=$& $(1/\sqrt{2}) \big( \ket{+00++}+\ket{+01-+}$  & $ X_{2}Z_{1}Z_{3}Z_{5},$ \\ 
 & & & $+ \ket{-10+-}+ \ket{-11--} \big)$  & $X_{3}Z_{2}Z_{4},$    \\
 & & &  & $ X_{4}Z_{3},$ \\
 & & &  & $X_{5}Z_{2}$  \\ 
 & & & & \\ \hline
\end{tabular}}
\end{table}

\subsubsection{Graph states as stabilizer states}

As we have seen in section \ref{sec:GKT}, any stabilizer state to which a Clifford group operator is applied remains a stabilizer state. The constructive definition defines a graph state as the result of just such a process, and so graph states are stabilizer states. As the initial states ($\ket{+}_i$) to which the $CZ$ operators are applied are stabilized by $X_i$, by evolving the generators of the stabilizer as discussed in section \ref{stab-evo}, we obtain the corresponding operators for the graph state. This yields an alternate definition for the graph state corresponding to some graph $G$ as the state for which the set of operators
\begin{equation}
S = {X_i \prod_{j \in N_G(i)} Z_j} 
\end{equation}
generate the stabilizer. Here $N_G(i)$ denotes the set of vertices connected by an edge to vertex $i$, which is called the neighborhood of vertex $i$.

So, all graph states are stabilizer states. What about the converse statement: are all stabilizer states also graph states?  Consider the two-qubit state stabilized by $X_{1}X_{2}$ and $Z_{1}Z_{2}$; this is an example of a stabilizer state that cannot be generated by the constructive definition, and so the converse statement is not true.  However, it can be transformed into a constructively defined two-qubit linear chain by the application, to either qubit, of a Hadamard gate. Since the Hadamard is a local Clifford operation, there exists a graph state with the same entanglement properties as our example stabilizer state. Surprisingly, this is always the case: all stabilizer states are local Clifford equivalent to some graph state\cite{VdNDDM01a,VdNDDM02a}.  A loose way of understanding this result is to consider how the constructive definition provides the $CZ_{i}^{j}$ gates, and so only the local gates $H_{i}$ and $P_{i}$ are required to complete the generators Clifford group. However, this mapping is not necessarily one-to-one, as shown in table~{\ref{EqGraphs}}, a single stabilizer state may be equivalent up to local unitary operations to many different graphs.

\begin{table}
 \tbl{Example stabilizer states with different equivalent graph representations.  Vertices labeled with operators are equivalent to the stabilizer state when these local unitaries are applied to the vertex after having followed the constructive prescription.\label{EqGraphs}} { \normalsize
\begin{tabular}{| c r | c c c | } \hline
 & & & &   \\ 
Equivalent graph  &  & &  Stabilizer & \\ 
representations  &  & &  generators & \\ 
\small
 & & & & \\ \hline 
 & & & & \\
 \multirow{2}{*}{\includegraphics{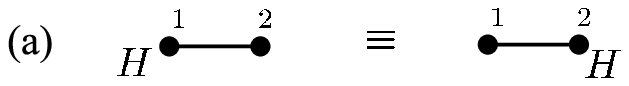} } 
& & & $X_{1}X_{2},$ & \\
& & & $Z_{1}Z_{2}$ & \\ 
& & & & \\ \hline
& & & & \\
 \multirow{3}{*}{\includegraphics{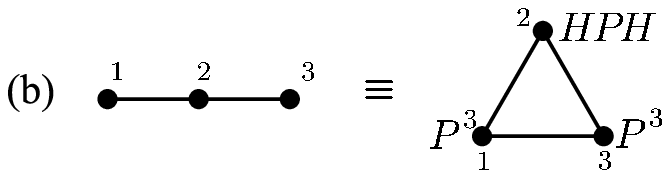} } 
& & & $X_{1}Z_{2},$ & \\
& & & $X_{2}Z_{1}Z_{3},$ & \\
& & & $X_{3}Z_{2}$ & \\ 
 & & & & \\
  & & & & \\ \hline												
\end{tabular}}
\end{table}

In table~{\ref{EqGraphs}} the required local unitaries have been explicitly added so that the different representations are strictly identical.  However, if we are interested in the required entanglement resource then local unitaries are
unimportant, and are comparatively easy to implement.  Therefore, different topology graphs can be sorted into local unitary (LU) equivalence classes.  Hein \textit{et al} have characterized all of the equivalence classes for graphs with up to 7 vertices\cite{HEB01a}.  They found that out of 995 graphs there are only 45 classes of LU-equivalent graphs.  Local Clifford (LC) equivalence is a special case of LU-equivalence, in which equivalence is only considered up to operations in the Clifford group. Van den Nest \textit{et al}\cite{VdNDDM01a} showed that the problem of identifying whether two graphs are LC-equivalent is reducible to a purely graphical problem;  two graphs are LC-equivalent if and only if there exists a series of \textit{local complementations} that transforms one graph into the other.  A local complementation is an operation that can be applied to any vertex $i$.  When applied to vertex $i$, every neighbour of $i$ gains a graph edge with every other neighbour.  However, since $(CZ^{i}_{j})^{2}=\id$, if the neighbours were already connected, the two graph edges cancel each other.  In table~{\ref{EqGraphs}b} we transform between the two graphs by performing a local complementation on vertex 2.  As vertex 2 has 1 and 3 as neighbouring vertices, the action of complementation connects or disconnects 1 and 3.  This leaves the problem of finding an algorithm to work through all possible combinations of local complementations, so as to determine whether two graphs are LC-equivalent.  Van den Nest \textit{et al}\cite{VdNDDM02a} showed that an algorithm developed by Bouchet\cite{Bo01a} --- initially for purely theoretical purposes --- will efficiently solve this problem in polynomial time.  
This algorithm proves extremely useful when planning how best to implement an algorithm using graph states. Many architectures do not allow for many gates to be simultaneously executed on a given qubit, meaning that edges must be created sequentially, as is the case for many schemes based on optical measurements, such as double-heralding\cite{BK01a} and repeat-until-success\cite{LBK01a}. In these cases it is sensible to cycle through LC-equivalent graphs to minimize the number of entangling operations necessary to construct the state.

\subsubsection{Pauli measurements on graph states\label{sec:graph-rules}}

\begin{table}[h!]
\label{tab:graph-rules}
  \tbl{Transformation rules for graph states undergoing local measurements of Pauli group operators. These describe the resultant state up to local Clifford operations. Here we adopt the convention of denoting sets by uppercase letters and elements of a set using lowercase letters. The equations in the table use $E(A,B)$ to denote the set of edges between sets $A$ and $B$, $E(A,B) = \{\{a,b\} \in E: a \in A, b \in B, a \neq b \}$, and the operation $E \Delta F$ is defined as $E \Delta F = (E \cup F) - (E \cap F)$. In the examples, vertex $a$ is highlighted in white.}   {\normalsize
\begin{tabular}{| l | l | } \hline

Rule &  Example  \\ \hline 
  &   \\
  \textbf{For a $Z$-basis measurement} of vertex a: & \multirow{7}{*}{\includegraphics[height=3cm]{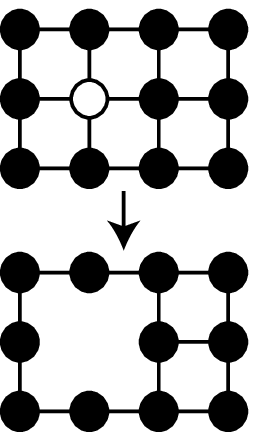} }  \\
 (1) Remove vertex $a$ and edges from $a$.  &  \\
 & \\ 
 Formally: $G'= G -\{a\}$&   \\
   &   \\ 
   &  \\
   &  \\
   &  \\ \hline
   &  \\
  \textbf{For a $Y$-basis measurement} of vertex $a$:   & \multirow{7}{*}{\includegraphics[height=3cm]{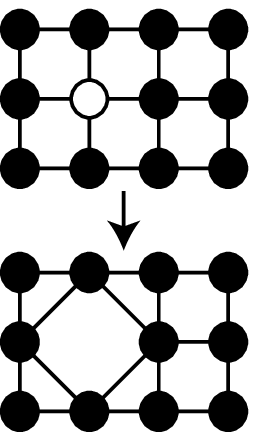} }  \\
 (1) Local complement vertex $a$;&   \\ 
  (2) Remove vertex $a$ and edges from $a$. &   \\
    &   \\
   Formally: $G'= G\Delta E (N_a,N_a) - \{a\}$ &  \\
     &   \\
     &  \\
     &  \\  \hline			
   &  \\
      \textbf{For an $X$-basis measurement} of vertex $a$, first choose a & \\  
      vertex $b$ (grey) that is a neighbor of $a$, and then:&  \\
      (1) Complement the neighbors of $b$ with the  neighbors of $a$; &  \multirow{7}{*}{\includegraphics[height=3cm]{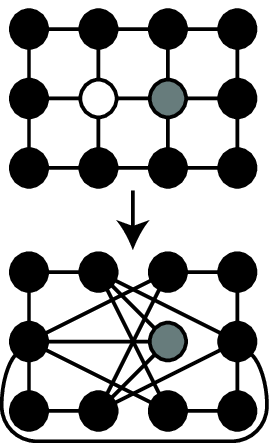} }  \\
      (2) Complement the mutual neighbors of vertices $a$ and $b$;&  \\
   (3) Complement vertex $b$ with the neighbors (except $b$) of $a$;& \\
     (4) Remove vertex $a$ and edges from $a$.  & \\
     Note that, depending on which vertex is chosen as $b$, the & \\
     final graph will appear different.  However, we are only&\\ 
    concerned with graphs up-to local unitaries, and different  &\\ 
    choices of the  vertex $b$ lead to final graphs differing by & \\
    only local unitaries.  & \\
         &  \\
  Formally: $G'= G \Delta \,  E(N_{b},N_a) \Delta \,  E(N_{b}\cap N_a,N_{b}\cap N_a)$ &   \\
   $\Delta \, E(\{b\}, N_a - \{b\}) -\{a\}$ &   \\ 
 &   \\ \hline										
\end{tabular}}
\end{table}

When graph states undergo Pauli measurements the transformation obeys a set of graphical rules, which were discovered independently by Hein \textit{et al}\cite{HEB01a} and by Schlingemann\cite{S01a}.  Here we will only describe the rules, up-to local Clifford operation, which will mean that the outcome of the measurement is unimportant.  The rule for $Z$ measurements is rather simple, the measurement removes the corresponding vertices from the graph.  Since $X$ and $Y$ Pauli measurements differ from $Z$ by a local Clifford operation, we expect they begin with a set of local complementations before removing a vertex from the graph.  Given a graph $G$, the rules by which the structure of the graph is changed by local Pauli basis measurements of vertex $a$ is summarized in table \ref{tab:graph-rules}, where we use of the notation of Hein \textit{et al}\cite{HEB01a} to formalize the transformation. 

 \begin{figure}[t]
  \includegraphics[width=\columnwidth]{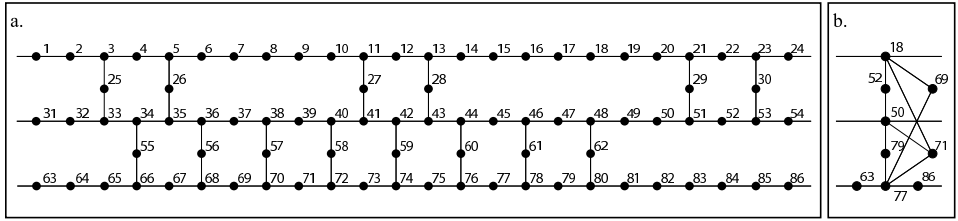}
  \caption{a) Cluster state fragment implementing a Toffoli gate. Qubits measured in the computational basis measurements have been removed. b) The minimal graph state obtained by removing the Clifford part of the first graph. The unmeasured qubits retain their index from (a). Removing the Clifford part of an algorithm in this way can dramatically reduce the number of qubits necessary to implement the computation. }
  \label{fig:minimalGraphstate}
\end{figure}
  
Our motivation for studying graph states is that measurements on them enable quantum computation, which we describe in the next section.  Since the effect of Pauli measurements can be easily simulated, it is interesting to ask what this might mean in terms of the underlying computation being performed.    Given the duality between Pauli measurements and Clifford group gates, the reader may not be surprised that Pauli measurements on a graph state execute the Clifford gates within a computation. The net result of this is that it is possible to remove the entire Clifford part of an algorithm, transitioning from a regular cluster state, to a more algorithm specific, less regular graph state. A graph from which the Clifford part has been removed is referred to as a \textit{minimal graph state}\cite{BES01a}, and an example of this removal process is shown in Fig.~\ref{fig:minimalGraphstate}. This ability to remove any Clifford part of an algorithm is strong motivation for considering general graphs, rather than remaining restricted to cluster states.

\subsection{One-way quantum computation\label{clustergates}}

In 2001, Raussendorf and Briegel proposed an innovative way to exploit the evolution of cluster states under measurement to perform universal quantum computation\cite{RB01a,RBB01a}. They proposed measurement patterns to perform arbitrary single qubit unitaries, and CNOTs between logical qubits. These patterns can then be superimposed on a regular square lattice cluster state. Qubits not used are simply measured out in the Z basis, effectively removing them from the graph. 

\subsubsection{Quantum teleportation}
Quantum teleportation\cite{BBCJPW01a} is not only an interesting application of quantum communication, but as we will see here, also offers a useful picture for understanding how quantum computation can be accomplished through carefully chosen measurements on a sufficiently large cluster state. While not as powerful as the stabilizer formalism described earlier, this approach is particularly intuitive and is useful for gaining some feeling for how these computations progress. A detailed explanation of this approach can be found in Refs.~\refcite{N01a},~\refcite{N01w}.

  \begin{figure}[h]
  \begin{center}
  \includegraphics[width=4cm]{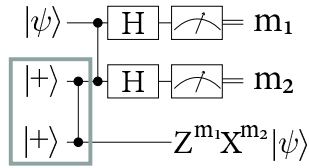}
     \end{center}
  \caption{Quantum teleportation of a single qubit. A quantum circuit used to prepare the entangled state is boxed in grey. The top two qubits are held by one party, say Alice, and the bottom qubit is held by a second party, say Bob.  Initially, Alice holds $\ket{\psi}$ and without using any quantum gates that interact with Bob, is able to transfer it to Bob up to some Pauli corrections.  Alice needs to transmit two classical bits of information for Bob to be able to apply the Pauli corrections and retrieve the teleported state.}
  \label{fig:teleportation}
    \end{figure}

Let us start by examining the teleportation protocol, which is best depicted as a circuit diagram in Fig.~\ref{fig:teleportation}. The steps are as follows:
\begin{enumerate}
\item A maximally entangled pair of qubits is generated, and shared between two parties.
\item The qubit to be transfered interacts with one half of the entangled pair by means of a controlled phase gate.
\item Both qubits held by the first party are then measured in the $X$ basis. 
\item The measurement results are communicated to the second party classically.
\item The second party performs a Pauli operator on their qubit dependent on the previous measurement results in order to obtain the teleported state.
\end{enumerate}

The quantum teleportation protocol can be seen as a concatenation of two iterations of the two-qubit teleportation illustrated below in Fig.~\ref{fig:teleportation2qubits}. This two-qubit teleportation procedure was proposed by Zhou \textit{et al}\cite{ZLC01a}.

 \begin{figure}[h]
  \begin{center}
  \includegraphics[width=4cm]{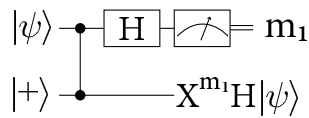}
 \end{center}
 \caption{The two-qubit quantum teleportation protocol.}
 \label{fig:teleportation2qubits}
 \end{figure}

We can follow this two qubit protocol mathematically. First, the entangling gate is performed between the qubit carrying the state to be teleported, $\ket{\psi} = \alpha \ket{0} + \beta \ket{1}$, and the second qubit prepared in the $\ket{+}$ state:
\begin{eqnarray}
CZ_{A}^{B}  \ket{\psi}_{A} \ket{+}_{B}=  \alpha \ket{0}\ket{+} + \beta \ket{1}\ket{-}.
\end{eqnarray}
We can emulate the effect of the X basis measurement, by instead performing a Hadamard rotation which will later be followed by a Z measurement:
\begin{eqnarray}
H_{A}CZ_{A}^{B}  \ket{\psi}_{A} \ket{+}_{B}&=& \frac{1}{\sqrt{2}} [\alpha (\ket{0,+} + \ket{1,+}) + \beta (\ket{0,-} - \ket{1,-})] \\
&=& \frac{1}{\sqrt{2}} [\ket{0} (\alpha \ket{+} + \beta \ket{-}) + \ket{1} (\alpha \ket{+} - \beta \ket{-})] \\
&=& \frac{1}{\sqrt{2}} (\ket{0}+\ket{1}X) (\alpha \ket{+} + \beta \ket{-}).
\end{eqnarray}
After a Z measurement with outcome $m_{1}$, the state is projected by $\bra{m_{1}}$:
\begin{eqnarray}
\bra{m_{1}}_{A}H_{A}CZ_{A}^{B}  \ket{\psi}_{A} \ket{+}_{B}  &=& X^{m_{1}} \left( \alpha \ket{+} + \beta \ket{-} \right) \\
&=& X^{m_1} H \ket{\psi}_{B}.
\end{eqnarray}
Here the teleportation protocol introduces a Hadamard gate onto the transmitted qubit. The evolution introduced in this way is the key to performing computation via measurement.

\subsubsection{Single qubit gates}

\begin{figure}[!h]
 \centering
 \includegraphics[width=0.85\columnwidth]{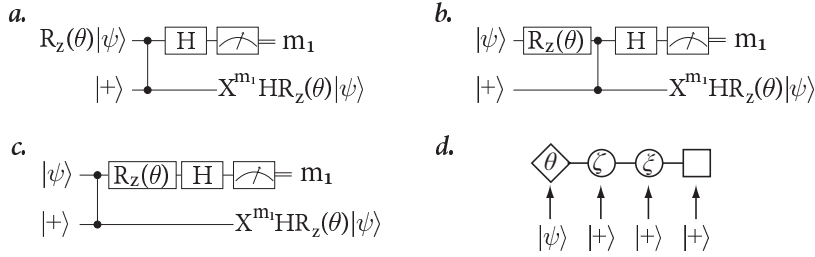}
  \caption{Rotation of a single qubit via measurements. a) The two qubit teleportation protocol. b) The rotation can be separated from the logical qubit. c) The $z$ rotation commutes with the CZ gate. d) The measurement pattern to  implement any single qubit gate exploiting the Euler angle formula, along with the initial state of each qubit. Here we use a diamond vertex to denote the input qubit and a square vertex to denote the output qubit.\label{fig:singlequbit}}
  \end{figure}

In the one-way model, single qubit gates are implemented by way of a variation of the two-qubit teleportation protocol: As any $Z$ rotations added to the state, $\ket{\psi}$, commute with the controlled phase gate and so can be absorbed into the measurement basis.  Rotating the $X$-measurement about the $Z$-axis enables a qubit measurement along any axis in the $X$-$Y$ plane. Fig.~\ref{fig:singlequbit} illustrates this process, showing three equivalent quantum circuits leading to the same output state.

By generalizing the measurement operator to allow for any $X-Y$ plane measurement basis, it is possible to apply an arbitrary operation of the form $J_ \theta = H e^{i \theta Z}$, for any real parameter $\theta$. This is equivalent to applying an arbitrary $Z$ rotation followed by a Hadamard gate. By concatenating three such teleportations, the resulting operator will by of the form: 
\begin{eqnarray}
J_\xi J_ \zeta J_ \theta & = & H e^{i \xi Z} H e^{i \zeta Z}  H e^{i \theta Z} \\ \nonumber
				             & = & H e^{i \xi Z} e^{i \zeta X} e^{i \theta Z}
\end{eqnarray}
Any single qubit unitary operation can be implemented in this way through suitable choice of $\theta $, $\zeta $ and $\xi$, a result which follows from the Euler angle decomposition of operators in \textit{SU}(2)\cite{NC01b}. The initial entangling operations used in this protocol can be seen as preparing a graph state, upon which single qubit measurements are made. As a result, measurements on a linear segment of a graph state can be used to apply any single qubit unitary operation, as depicted in Fig. \ref{fig:singlequbit}(d). 

\subsubsection{Entangling gates}
 \begin{figure}[!h]
\centering
  \includegraphics[width=0.65\columnwidth]{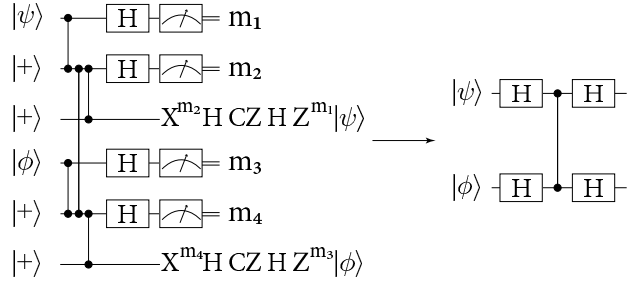}
 \caption{Entangling gate between two logical qubits via measurements.\label{fig:twoqubit}}
 \end{figure}
 
Just as single qubit rotations about the $z$ axis commute with the controlled phase gates used to construct the graph state, controlled phase gates also commute with each other, and so can simply be added to the graph to perform entangling gates between logical qubits. This yields another variant on the two-qubit teleportation protocol, which is shown in Fig. \ref{fig:twoqubit}.
 
\begin{figure}[!h]
\centering
 \includegraphics[width=0.5\columnwidth]{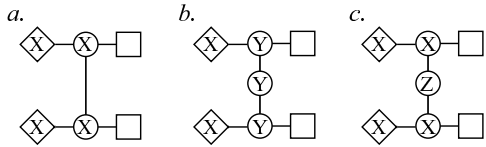}
 \caption{a) Measurement pattern for implementing the entangling gate shown in Fig.~\ref{fig:twoqubit}. This can be replaced with an arrangement of three qubits which allows the entangling gate to be performed using the measurement pattern in (b), or can be used to implement single qubit gates only, by using the measurement pattern in (c). As before, we use a diamond vertices to denote the input qubits and square vertices to denote output qubits.\label{fig:dynamic}}
 \end{figure}

This variation on the teleportation protocol is depicted in Fig.~\ref{fig:dynamic}a. A further variant on this approach is to add an additional node in place of the edge used to implement the controlled phase gate, as shown in Fig.~\ref{fig:dynamic}(b) and (c). This allows freedom of choice in the location of entangling gates, as measuring the intermediate qubit in the $Y$-basis recovers the entangling gate, while measuring it in the $Z$-basis leads to an identity operation. Conveniently, as the graph segments used for both single qubit gates and entangling gates are planar, and form a universal set of operations for quantum computation, it follows that any quantum circuit can be implemented by simply superimposing the measurement patterns corresponding to individual gates on a regular square cluster state. Additional qubits present in the original cluster state but not used in these gates can be removed simply by measurement in the $Z$-basis, as discussed in section \ref{sec:graph-rules}.
 
\section{Physical implementations}

We now turn our attention to how physical systems form a distributed architecture capable of producing entanglement between nodes.  We focus on hybrid matter-optical systems where matter systems embody the logical qubits and photons are used to generate entanglement between nodes.  The role of logical qubits is more suited to matter systems as matter storage is less problematic than photon storage.  However, photons can travel long distances without suffering significant decoherence from interactions with the environment.  Photons can be very robust\footnote{The degree of robustness depends on which photon degree of freedom is used to encode qubits.  For instance time-bin, spatial mode and frequency encoding are more robust than polarization encoding, as waveguide materials can cause polarization to drift.} to decoherence because in free space photons do not interact and when travelling through materials the degree of non-linearity is typically extremely weak.  However, generating entanglement by unitary evolution requires a non-linear Hamiltonian. Although there are some schemes for quantum computing that utilise weak non-linearities\cite{MKS01a}, this is not essential as non-linearities can also be induced by measurement.  Although there are a broad spectrum of optical methods for producing entanglement, we have in mind implementations that use only linear optics and non-number resolving photon detectors, as these present the lowest technological barriers.

We begin our introduction to the physics of these systems by introducing the physics of free photons and linear optical elements such as beam splitters, phase shifters, and wave-plates.  Next, we introduce atom-light interactions and how lasers can be used to manipulate matter qubits.

\subsection{Photons and linear optics}
\label{sec:Linear_Optics}

By quantization of the free space electromagnetic (EM) field within a cubic volume of space $V$, it is well known that the Hamiltonian is\cite{BR01b,BD01b}:
\begin{equation}
 	H_{\mathrm{EM}} = \sum_{\vec{\hat{e}} , \vec{k}} \left(  \frac{ k}{c} \right) \left(  a^{\dagger}_{\vec{k}, \vec{\hat{e}}} a_{\vec{k}, \vec{\hat{e}}} + \frac{1}{2} \right),
\end{equation}
where $a^{\dagger}_{\vec{k}, \vec{\hat{e}}}$ and $a_{k, \vec{\hat{e}}}$ are creation and annihilation operators for a photon of wavevector, $\vec{k}=k \vec{\hat{k}}$, and horizontal ($\vec{\hat{e}}=\vec{\hat{h}}$) or vertical ($\vec{\hat{e}}=\vec{\hat{v}}$) polarization\footnote{Note that, photon polarizations are always orthogonal to a photon's wavevector $\vec{\hat{e}}.\vec{k}=0$.}.  The sum over wavevectors covers all frequencies within the quantization volume and three independent Cartesian directions.  Note also that we take natural units where $\hbar=1$. It is useful to eliminate the need for the free space EM Hamiltonian by absorbing the time evolution into the photonic operators, so that $ a_{\vec{k}, \vec{\hat{e}}} \rightarrow a_{\vec{k}, \vec{\hat{e}}} \exp(-i k t/c)$.  Each photon corresponds to a plane wave oscillation of the electric field accompanied by the orthogonal magnetic oscillation. Conversely, the electric field operator for a point in space $\vec{r}$ at time $t$  is related to these photonic operators by a Fourier transform:
\begin{equation}
	\vec{E} (\vec{r}, t) = i \left( \frac{  k }{2 c \varepsilon_{0} V} \right)^{\frac{1}{2}} \sum_{\vec{k}, \vec{\hat{e}}}  \vec{\hat{e}} \left( a_{\vec{k}, \vec{\hat{e}}} \exp (-i k t/c + i \vec{k}. \vec{r}) - a_{\vec{k}, \vec{\hat{e}}}^{\dagger} \exp (i k t/c - i \vec{k}. \vec{r}) \right) ,
\end{equation}
where $\varepsilon_{0}$ is the permittivity of free space.  The state of electromagnetic (EM) field containing a single plane wave excitation is simply $a^{\dagger}_{\vec{k}, \vec{\hat{e}}} \ket{\mathrm{vac}}$ where $\ket{\mathrm{vac}}$ is the ground (vacuum) state of the EM field.  However, we will typically be dealing with localized single photons that are a superposition of plane waves:
\begin{equation}
	 a^{\dagger} = \sum_{\vec{k}, \vec{\hat{e}}} c(\vec{k}, \vec{\hat{e}})  a^{\dagger}_{\vec{k}, \vec{\hat{e}} },
\end{equation}
where normalization of the variable $c(\vec{k}, \vec{\hat{e}})$ ensures that $a^{\dagger}\ket{\mathrm{vac}}$ is an eigenstate of the number operator\footnote{The number operator counts the number of excitations and is $n= \sum_{\vec{k}, \vec{\hat{e}}}  a^{\dagger}_{\vec{k}, \vec{\hat{e}}} a_{\vec{k}, \vec{\hat{e}}}$} with eigenvalue $1$.  A localized photon is not an eigenstate of the free photon Hamiltonian and hence evolves in time non-trivially.  As we would expect this evolution results in propagation.  From the electric field operator we can see that this observable is the same for a later time $t'= t+ \Delta t$ at the shifted spatial point $\vec{r'}= \vec{r}+ t k / c$, which is consistent with a photon travelling across space at the speed of light.  

\begin{figure}
\centering
\begin{tabular}{c c}
\multirow{8}{*}{
\begin{tabular}{| l | c |} \hline
  Optical Element 					&  Photon operator transformation \\ \hline \hline
	Phase shifter($\phi$)   	&  $a \rightarrow \exp (i \phi) a $\\ \hline
	Waveplate($\theta, \phi$) &  $a_{h} \rightarrow \cos ( \theta ) a_{h} + \exp (i \phi) \sin (\theta) a_{v} $ \\ 
														&  $a_{v} \rightarrow \cos ( \theta ) a_{v} - \exp (i \phi) \sin (\theta) a_{h} $ \\ \hline
	Beam splitter             &  $a_{1} \rightarrow \cos ( \theta ) a_{3} + \exp (i \phi) \sin (\theta) a_{4} $ \\ 
  ($\theta, \phi$)          &  $a_{2} \rightarrow \cos ( \theta ) a_{4} - \exp (i \phi) \sin (\theta) a_{3} $ \\ \hline	
  Polarizing   							&  $a_{1, h} \rightarrow  a_{3, h} ;  a_{1, v} \rightarrow  a_{4, v} $ \\ 
	beam splitter							&  $a_{2, h} \rightarrow  a_{4, h} ;  a_{2, v} \rightarrow  -a_{3, v} $ \\ \hline						
\end{tabular}
} &

\multirow{8}{*}{ 
    \includegraphics{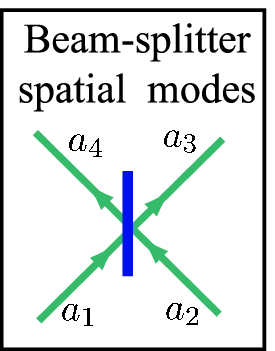}
} \vspace{3.5cm}
\end{tabular}
\caption{Some commonly used linear optical elements and the corresponding operator transformations.  Phase-shifters and waveplates act on a single spatial mode, though the latter has a polarization dependence.  Beam splitters and polarizing beam splitters map two input spatial modes, $a_{1}$ and $a_{2}$, to two output spatial modes,  $a_{3}$ and $a_{4}$.  For brevity the polarizing beam splitter is not completely general, but rather a perfect reflector of vertical light.  For the waveplate, the angle $\phi$ describes the relative dispersion of horizontal and vertical polarized light, whereas $\theta$ describes the dispersion of polarized light in a different basis.  For the beam splitter, $\phi$ describes the phase acquired passing through the beam splitter, and $\cos^{2}(\theta)$ is the percentage of transmitted (rather than reflected) light.\label{fig:linear_optics}}
\end{figure}

Having given a brief introduction to optics in free space, we now turn to the effects of linear optical elements on  localized single photons.  Linear optical effects are those where Heisenberg evolution of the creation operators is described by a mapping of vectors of operators $	\vec{a}^{\dagger}_{\mathrm{in}} = (a_{1}, a_{2}, ...a_{N})^{\dagger}$ that is unitary:
\begin{equation}
	\vec{a}^{\dagger}_{\mathrm{out}}	= U \vec{a}^{\dagger}_{\mathrm{in}}.
\end{equation}
Note that this criteria for linear optical unitaries excludes some more general unitary evolutions of the EM field.  Consider, spontaneous down-conversion where a single photon splits into two photons of lower energy $a^{\dagger}_{\vec{k}} \rightarrow a^{\dagger}_{\vec{k_{1}}}a^{\dagger}_{\vec{k_{2}}}$.  Although spontaneous down-conversion is part of a unitary process,  it is \textit{nonlinear} as the mapping of photon creation operators can take one creation operator to two creation operators.  In Fig.~{\ref{fig:linear_optics}}, we list some common linear optical elements and their effect on photonic operators.  All of these linear optical transformations can be produced by using materials of different reflectivity and refractive index, with polarization sensitive transformations requiring materials with polarization sensitive properties.  

In many circumstances linear optics produce results that are intuitive from a classical perspective.  However, one notably quantum effect is photon bunching, which occurs when otherwise identical photons impinge on different input ports of a $50/50$ beam splitter ($\theta=\pi/4$ and $\phi=0$), so that:
\begin{eqnarray} 
	a^{\dagger}_{1}a^{\dagger}_{2} \ket{\mathrm{vac}} & \rightarrow & \frac{1}{2} (	a^{\dagger}_{3} +  a^{\dagger}_{4})(	a^{\dagger}_{3} -  a^{\dagger}_{4}) \ket{\mathrm{vac}} ,\\ \nonumber
 & \rightarrow & \frac{1}{2} (	a^{\dagger}_{3} a^{\dagger}_{3}  - a^{\dagger}_{4} a^{\dagger}_{4}) \ket{\mathrm{vac}}.
\end{eqnarray}
Hence two identical photons tend to bunch together and leave a beam splitter from the same output port rather than following a classical distribution.  In any actual experiment photons are never perfectly identical, so the effect is typically measured by the Hong-Ou-Mandel dip\cite{HOM01a} that quantifies the deviation from classical behaviour.  

Later we shall see that bunching effects can be used to produce entanglement by enabling a partial measurement in the  Bell basis.  For a Bell state to be well defined there must be two different photonic modes (e.g. different polarizations or frequencies)  at each beam splitter input.  It is easy to verify that the singlet ($\ket{\Psi^{-}}$) is the only Bell state that can lead to unbunched photons:
\begin{equation}
	\frac{1}{\sqrt{2}}(a^{\dagger}_{1,h} a^{\dagger}_{2,v} - a^{\dagger}_{1,v} a^{\dagger}_{2,h})\ket{\mathrm{vac}} \rightarrow \frac{1}{\sqrt{2}}(a^{\dagger}_{4,h} a^{\dagger}_{3,v} - a^{\dagger}_{3,v} a^{\dagger}_{4,h})\ket{\mathrm{vac}}.
\end{equation}
This enables a \textit{partial} Bell measurement because detecting antibunched photons will distinguish the singlet from the other three Bell states; unlike a \textit{full} Bell state measurement that distinguishes all four Bell states from each other.

Since a partial Bell measurement introduces a probabilistic element to entanglement generation, a natural question is whether there is any way of performing a full Bell measurement with linear optics. However, Vaidman and Yoran\cite{VY01a} showed that linear optics cannot be used to perform measurements in a maximally entangled basis with probability $100 \% $. L\"{u}tkenhaus \textit{et al}\cite{LCS01a} extended this proof to cover schemes that use additional ancillas, forward feeding of measurement results or number-resolving detectors.  On the other hand, it is relatively easy to design apparatus to project onto the $\left\{ \vert 00 \rangle , \vert 11 \rangle , \vert 01 \rangle + \vert 10 \rangle, \vert 01 \rangle - \vert 10 \rangle \right\}$ basis\cite{BM01a,PYF01a,KLM01a,BPH01a,BK01a,PYF02a}.

\subsection{Matter-light interactions}
\label{sec:matter-light}

Since linear optical technology is most mature for optical wavelength photons, we will typically be interested in electronic transitions at optical wavelengths.  Within a matter system there may be several different available transitions, but here we consider the simplest possible system where an excited electronic state $\ket{e}$ decays to a ground state $\ket{1}$ via the emission of a photon.  If the purely EM contribution is removed from the Hamiltonian by absorbing the evolution into the photonic operators, then the remaining Hamiltonian for the system is:
\begin{equation}
	H = H_{\mathrm{matter}} + H_{\mathrm{int}},
\end{equation}
where $H_{\mathrm{matter}}$ is simply the difference in the energy, $ \omega$,  between $\ket{1}$ and $\ket{e}$:
\begin{equation}	
  H_{\mathrm{matter}} =  \omega \sigma^{+} \sigma^{-},
\end{equation}
which has been expressed in terms of the atomic raising and lowering operators, $\sigma^{+}=\kb{e}{1}$ and $\sigma^{-}=\kb{1}{e}$.  The term $H_{\mathrm{int}}$ is the interaction Hamiltonian between the two systems.  Typically the strongest matter-light interaction is the electric dipole transition\footnote{Transitions may be electric dipole disallowed, but may still occur by the weaker magnetic dipole or higher order electric contributions such as electric quadrupole.  Nevertheless,  the essential physics is captured by considering only electric dipole transitions.}, where the energy arises from the interaction of the electric dipole moment with the electric field:
\begin{equation}
            H_{int} = - \int \bm{\mu} (\vec{r}) \cdot \vec{E} (\vec{r}, t) d \vec{r},
\end{equation}
where $\bm{\mu}(\vec{r})$ is the electric dipole operator.  Since optical wavelengths are much longer than atomic distances, we can make the electric dipole approximation that assumes the electric field is constant over the whole volume where the dipole moment is nonvanishing. When the matter system is centered at the origin, this gives a simplified expression for the electric field that factors outside the integral:
\begin{equation}
            H_{int} = - i \sum_{\vec{k}, \vec{\hat{e}}}  \left( \frac{   |k| }{ \varepsilon_{0} V t} \right)^{\frac{1}{2}} \left( \int \bm{\mu} (\vec{r}) d \vec{r} \right) \cdot  \vec{\hat{e}} \left( a_{\vec{k}, \vec{\hat{e}}} \exp (-i |k|t/c ) - a_{\vec{k}, \vec{\hat{e}}}^{\dagger} \exp (i |k|t/c ) \right).
\end{equation}
The electric dipole operator is simply the distance of charge from the origin:
\begin{eqnarray}
	\bm{\mu}(\vec{r}) & = & -e \vec{r} , \\ \nonumber
										& = & -e ( x \vec{\hat{e_{1}}} +  y \vec{\hat{e_{2}}} + z \vec{\hat{e_{3}}} ) ,
\end{eqnarray}
where $e$ is the charge of an electron and $\vec{e_{i}}$ are unit vectors in Cartesian coordinates.  

Since we are only interested in the action of $\int \bm{\mu}$ on the $\{ \ket{1}, \ket{e} \}$ subspace, we rewrite the dipole operator in terms of transition elements: 
\begin{equation}
	\int \bm{\mu}(\vec{r}) d \vec{r} = \sum_{\vec{\hat{e_{i}}}} \sum_{a, b \in \{ 1, e\}}  \kb{a}{b} \vec{\hat{e_{i}}} \left( \int  \psi^{*}_{a}(\vec{r})  \psi_{b}(\vec{r}) \vec{r}.\vec{\hat{e_{i}}}  d \vec{r} \right)  ,
\end{equation}
and hence we require some knowledge of the electronic wavefunctions $\psi_{1}(\vec{r})$ and $\psi_{e}(\vec{r})$.  In the solid state the problem can be quite involved as an atom in a crystal lattice will not have a rotationally symmetric Hamiltonian, and hence will not have rotationally symmetric energy eigenstates.  

However, the essential physics is captured by considering the simple energy eigenstates of an atom in free space.  The charge of the atomic nucleus provides a rotationally symmetric Hamiltonian for the orbiting electron, whose energy eigenstates are charge balanced $|\psi_{i}(\vec{r})|=|\psi_{i}(\vec{-r})|$.  This symmetry entails that the following integrals vanish:
\begin{eqnarray}	
  \int  \psi^{*}_{1} (\vec{r}) \psi_{1} (\vec{r}) \vec{r}.\vec{\hat{e_{i}}}  d \vec{r} & = & 0 ; \forall i ; \\ \nonumber
  \int  \psi^{*}_{e} (\vec{r}) \psi_{e} (\vec{r}) \vec{r}.\vec{\hat{e_{i}}}  d \vec{r} & = & 0 ; \forall i ; 
\end{eqnarray}
so that the non-transitioning elements ($\kb{1}{1}$ and $\kb{e}{e}$) also vanish.  Further progress is easy if the ground state is an $S$-shell electronic state and the excited state is a $P$-shell electronic state with angular momentum $+1$ in the $\vec{\hat{e_{3}}}$ (z) direction\footnote{Rotational symmetry can be reintroduced by also accounting for the presence of the $P$-shell electronic state with $-1$ angular momentum.}.  For such a transition the symmetry of these states entails that the $\vec{\hat{e_{3}}}$ transition elements vanish, and the $\vec{\hat{e_{1}}}$ and $\vec{\hat{e_{2}}}$ elements only differ by a factor of $i$.  Therefore, for this simple system the electric dipole operator becomes:
\begin{equation}
	\int \bm{\mu}(\vec{r}) d \vec{r} = (\alpha \sigma^{+} + \alpha^{*}\sigma^{-} ) \vec{\hat{e_{1}}} + i(\alpha \sigma^{+} - \alpha^{*}\sigma^{-} ) \vec{\hat{e_{2}}} ,
\end{equation} 
where the variable $\alpha$ depends on further details of the electronic wavefunctions.  The phase of $\alpha$ is trivial as it follows from the global phase of the electronic wavefunctions, $\psi_{i}$ and $\psi_{j}$, and hence we take $\alpha=|\alpha|$. With this expression for the electric dipole operator the Hamiltonian is still too unwieldy to admit any analytic solutions.  However, the Hamiltonian is greatly simplified by neglecting terms that rotate quickly in the interaction picture\footnote{Where evolution due to $H_{EM}$ and $H_{\mathrm{matter}}$ have been absorbed in the the photonic and atomic operators, leaving $H_{\mathrm{int}}$ as the only remaining term in the Hamiltonian.}.  We have already brought the creation operators into the interaction picture, and absorbing the effect of $H_{\mathrm{matter}}$ into the matter-system operators gives $\sigma^{+} \rightarrow  \sigma^{+}  \exp^{i \omega t}$ and $\sigma^{-} \rightarrow \sigma^{-}  \exp^{- i \omega t}$.  The quickly rotating terms, $\sigma^{+} a^{\dagger } \exp^{i (\omega + k /c )t }  $ and $\sigma^{-} a \exp^{-i (\omega + k /c)t} $, have negligible impact on long term evolution.  Neglecting them is the so-called \textit{rotating wave approximation} and gives the interaction Hamiltonian:
\begin{eqnarray}
\label{eqn:Ham_Rot_Wav}
		H_{\mathrm{int}} & = & -i W_{k}  \left( \sum_{k}   \sigma^{+} b_{k} \exp (- i \Delta_{k} t  )	- \sigma^{-} b_{k}^{\dagger} \exp (i \Delta_{k} t  )  \right) ,
\end{eqnarray}
where $\Delta_{k}=(\omega - k/c )$ is the detuning,  $W_{k}$ is the coupling strength:
\begin{equation}
	W_{k} = 2 \alpha \left( \frac{   |k| }{ \varepsilon_{0} V c t} \right)^{\frac{1}{2}} ,
\end{equation}
and the new photonic operator is:
\begin{equation}
  b_{k} = \frac{1}{2} \left( a_{k \vec{e_{3}}, \vec{e_{1}}} + ia_{k \vec{e_{3}}, \vec{e_{2}}}   +  a_{k \vec{e_{2}}, \vec{e_{1}}} + i a_{k \vec{e_{1}}, \vec{e_{2}}}    \right).
\end{equation}
This Hamiltonian shows that atomic excitation absorbs a photon, and that the reverse process of atomic relaxation always emits a photon.  Note also that these processes conserve angular momentum, as the photonic operators $b_{k}$ are  $+1$ angular momentum eigenstates\footnote{Angular momentum conservation is a consequence of the invariance of physics under rotations of the reference frame.  Similarly, angular momentum eigenstates are rotationally symmetric up to a phase.  For an eigenstate of eigenvalue $m$, the eigenstate acquires a phase of $\exp (i m \theta \pi)$ for a rotation of $\theta$ degrees about the angular momentum axis.  Notice that the $b_{k}$ operators satisfy this rotational symmetry}, so when emission occurs the photons carry away the angular momentum that is lost when the electron transitions from a $P$ to $S$ electronic state.  

In deriving this Hamiltonian we have made three assumptions: (\textit{i}) The electric field is uniform over the matter system;  (\textit{ii}) rapidly oscillating terms can be neglected; (\textit{iii}) the electronic wavefunctions are free space atomic wavefunctions.  The last assumption essentially gave us the polarization features of our Hamiltonian, and for matter qubits embedded in a crystal lattice we can expect a similar result except with different polarization features.  For example, in both quantum dots and nitrogen-vacancy centres in diamond (NV centres) the $X-Y$ plane rotational symmetry can be broken and there are often two separate energy eigenstates that decay via horizontal and vertical polarized light\cite{DH01a,GSSKP01a,PRWH01a,KAKH01a,SFSFBGODRRGRJP01a}. In NV centres the crystal lattice\footnote{The NV defect has trigonal discrete symmetry, which breaks the continuous rotational symmetry of a free atom in space.  However, the effects of symmetry breaking are amplified by strain inside the crystal lattice.} breaks the symmetry, whereas in quantum dots the dominant effect is the elliptical shape of typical dots.

Having derived Hamiltonian Eqn.({\ref{eqn:Ham_Rot_Wav}}), we now need to consider how to solve the Schr\"{o}dinger equation for systems evolving under this Hamiltonian.  In general, this is a difficult problem.  However, by assuming that the coupling strength is approximately constant over the region where the detuning is small, $W_{k} \sim W_{\omega} $,  one can derive Weisskopf-Wigner decay\cite{BR01a,Ved01b}.  This analytic solution gives an exponential decay of the excited state at a rate of $2 \pi W_{\omega}$, and will also predict many spectral features of the resulting photon.  

However many proposals for a mature hybrid technology employ the cavity effects of quantum electrodynamics, which occur between reflecting surfaces.  In this context the constant coupling strength assumption is far from true.  These cavities dramatically affect the coupling strengths of different frequency photons, depending on whether the frequencies are resonant with the cavity geometry. The construction of such a cavity can take many forms, including  spherical mirror cavities~\cite{Keller03,Wilk07,Kimble08}, micropillars\cite{SSBLPCB01a}, microdisks\cite{KMBGIZHSP01a}, microtoroidal resonators\cite{Aoki06} and photonic crystals\cite{HTKRKF01a,YSHKGRESD01a,BHADHPI01a}.  Interesting phenomena occur when a matter qubit is placed within a cavity, and under certain very general conditions is described by the Jaynes-Cummings model\cite{JC01a}.  In this model we assume that one resonant cavity frequency dominates evolution and that all other frequencies can be neglected.  Assuming that this frequency is on resonance with the electronic transition, we have:
\begin{equation}
	H_{\mathrm{JC}} = i g_{\omega}  ( \sigma^{-} b^{\dagger}_{w}  - \sigma^{+} b_{w}  ),	
\end{equation}
where the photon $b^{\dagger}_{w}$ only extends across the volume of the cavity, and hence the relevant quantization volume is that of the cavity; we use $g_{\omega}$ rather than $W_{\omega}$ to identify variables with dependence on cavity volume.  Furthermore, the photon is now a standing wave rather than a plane wave as the cavity walls enforce boundary conditions on the EM field.  Consequently, the coupling strength is modified by a factor of $\sin (\omega x/c) $, where $x$ is the distance of the matter system from either cavity wall\footnote{This assumes a 1D model of the cavity, though a full 3D model will include a similar sinusoidally varying effect on coupling strength.}.

Important effects of this Hamiltonian include Rabi oscillations and the Purcell effect.  Rabi oscillations occur when the system is initialized in a state with a single excitation of either matter qubit or cavity mode, and they correspond to oscillations of the excitation between the electron and cavity.  When the system is not lossy, the Rabi oscillations complete full cycles with a frequency $ g_{w}$.  When the system is lossy, the Rabi frequency increases and stimulated emission from the cavity is enhanced (a phenomena known as the Purcell effect\cite{PTP01a,Pur01a}).

This introduction of the Jaynes-Cummings model has been idealized as we have assumed that photons do not escape from the cavity or get absorbed by the cavity walls.  The quality factor of a cavity, $Q$, is a measure of that cavities' performance at retaining a quanta of energy.  The value of $Q$ is inversely proportional to the probability, per Rabi cycle, that the excitation is lost from the system.

For the purposes of quantum computation, photons lost by absorption into the cavity walls are undesirable.   However, leakage of photons into modes outside the cavity can be desirable, provided that the leakage is into one particular, chosen, mode.  This can be achieved by thinning a region of one reflecting surface, which increases the probability of tunnelling via this route.

For DQC applications we will need our cavity to leak photons into a mode that is monitored with photon detectors.  If we monitor a joint matter-cavity system for the emission of photons, and none are detected, then its evolution can be described  by a non-Hermitian conditional Hamiltonian\cite{GZ01b}:
\begin{equation}
\label{eqn:Ham_Leaky}
	H_{\mathrm{leaky JC}}= H_{JC}  - \frac{i }{2} J^{\dagger} J,
\end{equation}
where $J$ is known as the quantum jump operator,
\begin{equation}	
	  J=\sqrt{\kappa} b_{\omega} \; ,     
\end{equation}
with $\kappa$ quantifying the leakage rate of the cavity.  It is the non-Hermitivity of the additional term that is responsible for the irreversible\footnote{Here we use irreversible to mean not cyclic, in contrast to Rabi oscillations.  When the Jaynes-Cummings approximation is not made, we observe this kind of irreversibility in free-space photon emission.  In the free-space case, irreversibility occurs because we couple to an infinite number of field modes, and hence any oscillation would take an infinite amount of time.  Since, the Jaynes-Cummings model involves a finite number of modes, this irreversibility must be added in by hand.} evolution of the system.  This irreversible decay also reduces the norm of the wavefunction $N= \langle \Psi \vert \Psi \rangle$ at a rate $\dot{N}$ that represents the probability density of detecting a photon.  In general, any measurement event associated with a jump operator $J$ will occur with probability $\langle \Psi \vert J^{\dagger} J \vert \Psi \rangle$.  If a detection event occurs, the system is projected by the jump operator.

\subsection{Local rotations}

In this review we are interested in how entanglement can be produced between different nodes of a distributed quantum computer. However, this presupposes that we already have excellent control over the quantum systems within a node.  Indeed, we will regularly assume that we can perform arbitrary local operations to high fidelity, with imperfections in local operations analysed as small noise contributions.

Different physical systems allow different mechanisms for performing local unitaries.  For example, when we wish to rotate between a ground state $\ket{1}$ and an optically accessible excited state $\ket{e}$, this can be accomplished by applying a laser at a frequency resonant with the transition frequency.  The angle $\theta$ of the rotation, e.g. $\ket{1} \rightarrow \cos(\theta) \ket{1} + \sin (\theta) \ket{e}$, is determined by the laser amplitude integrated across the duration of the pulse.  Of particular interest are so-called $\pi$ pulses that pump $\ket{1} \rightarrow \ket{e}$.  Note also, that if each node contains more than a single qubit, then arbitrary rotations between those qubits are also considered to be local. For example, a control NOT gate between an electron spin and a near-by nuclear spin is local to the node and can also be implemented optically\cite{DCJTMJZHL01a,NMRHWYJGJW01a}. In contrast, gates between qubits in different nodes can not be implemented so easily, and require an entanglement generation protocol.

Since we assume that local gates are technologically easy compared to non-local gates, the curious reader may wonder why bother with distributed quantum computing? Surely we could build a quantum computer using these techniques for local gates? Unfortunately this is often not the case. Arbitrary control over a few qubits at each node is much more technologically viable than arbitrary control of an arbitrary number of qubits.  If we try and scale up nodes to larger sizes, numerous technological difficulties emerge.  One of these is the problem of \textit{addressability} in completely local quantum computers, which occurs when we only want one qubit to interact with a particular control field\cite{Seth01a}.  This can not be performed by spatial resolution as interacting qubit systems will typically be separated by less than an optical wavelength.  As for frequency resolution, this may be possible for a limited number of qubits. For example, in a 2 qubit system frequency splitting may occur because the qubits are of different species, such as electron and nuclear spin.  However, in long chains of identical qubits there will be no effect that splits the frequency of every qubit by a resolvable amount. 

Although addressability is an obstacle to large-scale local architectures, it is certainly not an insurmountable obstacle.  For example, there exist so-called \textit{global control} schemes that circumvent the addressability problem\cite{Seth01a,Benj01a,fitzsimons-2005a,Love01a,FXBJ01a}. Even when addressability can be achieved, cooling presents an additional hurdle, as it is often extremely difficult to reduce population of low lying excited states in such systems.  

Aside from the question of technological viability, distributed quantum computers have other inherent advantages.  Typically, a distributed architecture can be made fault tolerant at higher noise levels than a local architectures\cite{Svore2005,Svore2006}.  Furthermore, distributed architectures  allowing for arbitrary connectivity offer computational performance superior to local architectures\cite{VanMeter06}.  As such, it may prove advantageous to develop a distributed architecture even if a local architecture is viable.  Having outlined some issues involved in local operations, and local architectures, we return to our main concern: how to generate entanglement between local nodes. 

\subsection{Single-photon protocols}
\label{sec:single_photon_protocols}

Cabrillo \textit{et al}\cite{CCFZ01a} proposed the first protocol for generating entanglement between distributed atoms via photon measurements, and this seminal work has spawned numerous spin-off protocols.  Bose \textit{et al}\cite{BKPV01a} showed that this protocol could be adapted for the purpose of implementing teleportation.  Browne \textit{et al}\cite{BPH01a} used adiabatic laser pulses to improve the protocol's success probability and robustness to various errors\footnote{For example, by using detuned lasers they can filter out scattered laser light.}.  Childress \textit{et al}\cite{CTSL01a} demonstrated that a different electronic level structure can provide the same benefits.  Here we will outline the essential components of the Cabrillo proposal.  For the purpose of better comparison with later proposals we will assume that two photon detectors are used\footnote{In the original proposal only a single photon detector was proposed.  This wastes an output port of the beam splitter and reduces the success probability by $1/2$.}. 

\begin{figure}[t]
\centering
\includegraphics{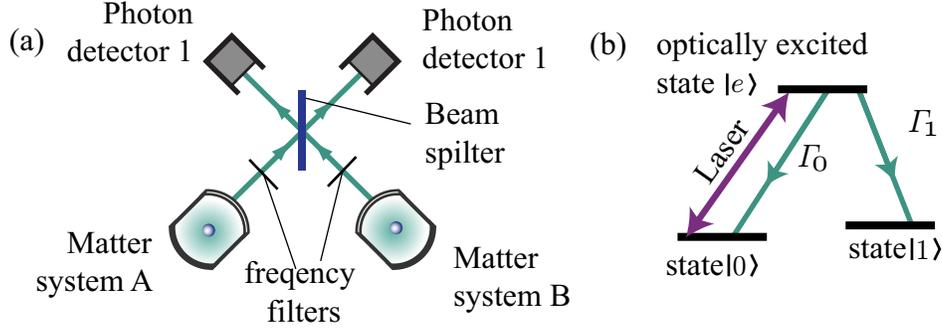}
\caption{single-photon protocols: (a) The experimental set-up used for single-photon protocols (and many other protocols).  Two matter systems embodying separated qubits are placed so that emitted photons are encouraged to emit towards a beam splitter, this can be achieved by using QED cavities or collection lens.  Before the beam splitter are frequency filters that absorb light emitted from the $\ket{e}\rightarrow \ket{0}$ transition.  After the beam splitter are two photon detectors. (b) The $\Lambda$-level structure required for each matter system in the original Cabrillo proposal.}
\label{fig:Cabrillo}
\end{figure}

Cabrillo \textit{et al} considered two matter systems with a $\Lambda$-level structure, with an excited state $\ket{e}$ that decays to ground states $\ket{0}$ and $\ket{1}$ with decay rates $\Gamma_{0}$ and $\Gamma_{1}$ respectively (see Fig.~{\ref{fig:Cabrillo}b}).  These matter systems are arranged in the experimental set-up shown in Fig.~{\ref{fig:Cabrillo}a}.  Note that frequency filters ensure that only photons from the $\ket{e} \rightarrow \ket{1}$ can reach the detectors.  The Cabrillo proposal begins with both matter systems prepared in the state $\ket{0}$.  Next a short pulse of laser light resonant with the $\ket{0} \leftrightarrow \ket{e}$ transition is used to pump a small proportion of the population into the state $\ket{e}$, giving the joint state:
\begin{equation}
 ( \cos(\theta) \ket{0}_{A} + \sin(\theta) \ket{e}_{A} )( \cos(\theta) \ket{0}_{B} + \sin(\theta) \ket{e}_{B} ) ,
\end{equation}
where $\theta$ is a small parameter that determines the strength of the weak excitation.  Since there is only a small probability of excitation, usually no photons will be detected.  However, when a photon is detected we project out the $\ket{0}_{A} \ket{0}_{B}$ component.  The purpose of choosing  small $\theta$ is so we can assume that two photon processes are negligible, so $\ket{e}_{A} \ket{e}_{B}$ also vanishes.  This is necessary because although only one photon was detected, photon loss means that more could have been emitted.  Later we derive the infidelity due to non-zero $\theta$, though for now we assume the two photon term is negligible.   The presence of the beam splitter ensures that we do not know where the emission originated, and hence we are in a maximally entangled Bell state:
\begin{equation}
 \ket{\Psi^{\phi}} = \frac{1}{\sqrt{2}} \left( \ket{0}\ket{1} +  e^{i \phi} \ket{1}\ket{0}   \right),
\end{equation}
where the variable $\phi$ depends on the path lengths to the beam splitter and which detector clicked.  All single photon schemes require that the apparatus is interferometrically stable to distances much shorter than an optical wavelength so that the phase factor $\phi$ is known.

Now we account for the infidelity due to non-zero $\theta$, which corresponds to double excitation processes.  If two photons are emitted, and one is undetected, it will eventually be absorbed by the environment.  Equivalently, both may reach a given detector which is unable to resolve, or count, the pair.  Either process is described by a measurement of the number of photons, but without our knowledge of whether the answer was 1 or 2.  Hence, we have an incoherent mixture of the desired state with $\ket{11}$:
\begin{equation}
 (1- \eta) \kb{\Psi^{\phi}}{\Psi^{\phi}} + \eta \kb{11}{11} ,
\end{equation}
where $\eta$ depends on the excitation strength and photon loss rate, and for simplicity we have assumed that $\Gamma_{0} \ll \Gamma_{1}$.  The ratio of the weighting $(1-\eta):\eta$ is proportional to the relative probability of one and two photons resulting in a single detector click.  Defining $T$ as the probability of transmission and detection of a given photon emitted by either matter system, it follows that one photon emission and detection occurs with probability:
\begin{equation}
	P_{(1)} = 2 \sin^{2} (\theta) \cos^{2} (\theta) T.
\end{equation}
The equivalent probability for two photon emission followed by one detector click is:
\begin{equation}
	P_{(2)} = \sin^{4} (\theta) (1- (1-T)^{2}) ,
\end{equation}
where we have assumed that the photons bunch and detectors can not resolve the number of photons.  It follows that the erroneous contribution has magnitude:
\begin{equation}
  \eta = \frac{P_{(2)}}{P_{(1)}+P_{(2)}}  = \frac{\sin^{2}(\theta) (2- T)}{2 - \sin^{2}(\theta)T}.
\end{equation}
To reduce $\eta$ we must reduce $\theta$, and hence the probability of detecting a photon.  In the small $\theta$ and small $T$ regime, we have direct proportionally between infidelity and success probability: $\eta \sim \theta^{2}$ and $P_{(1)}+P_{(2)} \sim 2 \theta^{2}$.

Note that in Cabrillo \textit{et al}'s original proposal the decay channel $\Gamma_{0}$ only reduces the success probability.  Therefore, we can also use a similar approach to entanglement generation where only the state $\ket{1}$ couples to $\ket{e}$.  For such a scheme we prepare the superposition $\cos(\theta)\ket{0} + \sin(\theta)\ket{1}$ and then $\pi$ pump $\ket{1} \rightarrow \ket{e}$.    

Entanglement from a single photon detection has been demonstrated between two clouds of cold caesium atoms\cite{DCZ01a,CRFPEK01a}.  In these experiments the logical states of the qubits were the: $\ket{0}$ all atoms in ground state; $\ket{1}$ an ensemble of atoms sharing an excitation.  

\subsection{Two photon protocols}
\label{sec:two_photon_protocols}

Unlike single-photon schemes, the later two-photon schemes are robust against interferometric instability and do not have their success probability tied to infidelity.  As outlined earlier we will introduce a typical anti-bunching protocol and the double-heralding protocol.  It is a testament to the promise of distributed quantum computing that an anti-bunching scheme has successfully been used to generate entanglement between two remote trapped ions\cite{MMOYMDM1a,MMOYMM01a}.  

The repeat-until-success (RUS) protocol\cite{LBK01a,LBBKK01a} is another interesting protocol that we do not have space to review.  However, it is worth noting that the RUS protocol performs a generalized measurement\footnote{By generalized measurement we mean Positive Operator Values Measurement, or POVM for short} with outcomes that are locally equivalent to the identity or control-$Z$ gates.  Therefore, it is the only known hybrid protocol that has no failure\footnote{The identity operation is not a successful entangling gate.  However, if the qubits hold some pre-existing entanglement with third parties, this third-party entanglement is not destroyed by the identity gate.  In this context, the identity gate can be interpreted as neither a success nor a failure.} outcome when there is no photon loss and detectors can resolve different numbers of photons.

\subsubsection{Bunching based schemes}
\label{sec:Anti_Bunch}
 
Two identical photons impinging on different different ports of a beam splitter will bunch together and leave through the same beam splitter port (see section {\ref{sec:Linear_Optics}}).  This feature of Bell states has been exploited in numerous proposals for generating entanglement either using polarization\cite{FZLGX01a,DK01a} or frequency\cite{DMMMKM01a} to define the logical basis.  All of these schemes essentially use a $\Lambda$-level structure and success is heralded by anti-bunched photon detections, from which we can infer that the photons were \textit{not} identical. Here we will describe the proposal\footnote{We give their more refined version of the protocol that uses more experimental apparatus but increases the probability of success.} of Duan and Kimble\cite{DK01a}.

\begin{figure}[t]
\centering
\includegraphics{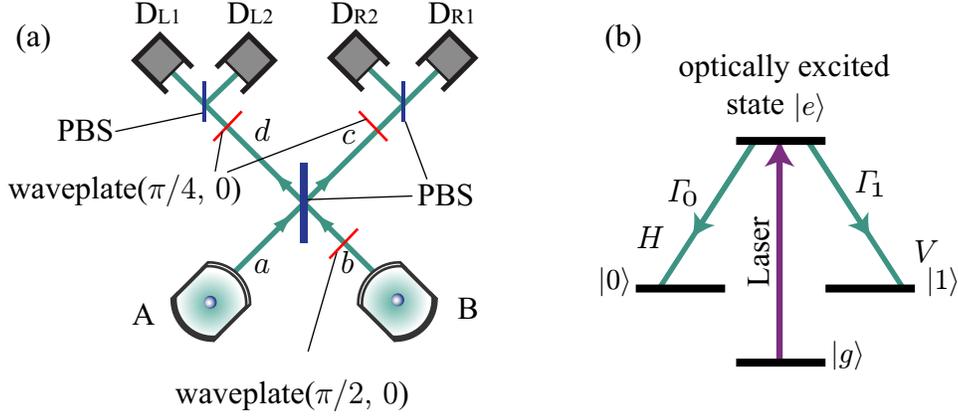}
\caption{An outline of Duan and Kimble's proposal for generating entanglement that is heralded by detecting anti-bunched photons. (a) the experimental apparatus includes 3 waveplates, 3 polarizing  beam splitters (PBS) and four photon detectors.  $D_{L1}$ and $D_{L2}$ are the ``left'' detectors and the ``right'' detectors are $D_{R1}$ and $D_{R2}$.  (b) the electronic level structure.}
\label{fig:AntiBunch_Outline}
\end{figure}

Duan and Kimble proposed an experimental set-up that uses waveplates and polarizing beam splitters to erase which-path information of polarization encoded photons (see Fig.~{\ref{fig:AntiBunch_Outline}}).  Their matter system has a level structure with an excited state that decays to $\ket{0}$ or $\ket{1}$ via the emission for a horizontal or vertical photon, respectively.  They also assume a ground state $\ket{g}$ that can be driven to the state $\ket{e}$.  The role of $\ket{g}$ is only as a storage state before driving the state to $\ket{e}$, so it is unnecessary if we have some other means of preparing $\ket{e}$.  The protocol runs as follows:
\begin{enumerate}
	\item From the initial state $\ket{g}\ket{g}$, adiabatically drive both matter-cavity systems into the state $\ket{e}\ket{e}$;  
	\item wait for the system to relax whilst monitoring for emitted photons; the protocol succeeds when we observe two photon detection events, one in a left detector and one in a right detector.
\end{enumerate}
Just after the matter systems $A$ and $B$ relax and emit a photon, the state of the system is:
\begin{equation}
	\ket{\psi} = ( g_{0} \ket{0}a^{\dagger}_{\vec{\hat{h}}}  + g_{1} \ket{1}a^{\dagger}_{\vec{\hat{v}}})( g_{0} \ket{0}b^{\dagger}_{\vec{\hat{h}}}  + g_{1} \ket{1}b^{\dagger}_{\vec{\hat{v}}}) \ket{\mathrm{vac}},
\end{equation}
where $g_{0}$ and $g_{1}$ are the relative probabilities of relaxation by horizontal versus vertically polarized photons.  The first waveplate rotated the polarization at mode $b$:
\begin{equation}
	\ket{\psi} = ( g_{0} \ket{0}a^{\dagger}_{\vec{\hat{h}}}  + g_{1} \ket{1}a^{\dagger}_{\vec{\hat{v}}})( g_{0} \ket{0}b^{\dagger}_{\vec{\hat{v}}}  - g_{1} \ket{1}b^{\dagger}_{\vec{\hat{h}}}) \ket{\mathrm{vac}}.
\end{equation}
Impacting on the PBS the vertical photons are reflected:
\begin{equation}
	\ket{\psi} = (-g_{0}^{2}\ket{00}c^{\dagger}_{\vec{\hat{h}}}c^{\dagger}_{\vec{\hat{v}}} - g_{0}g_{1} \ket{01} c^{\dagger}_{\vec{\hat{h}}}d^{\dagger}_{\vec{\hat{h}}} +  g_{1}g_{0}\ket{10}d^{\dagger}_{\vec{\hat{v}}}c^{\dagger}_{\vec{\hat{v}}} + g_{1}^{2}\ket{11}d_{\vec{\hat{v}}}d^{\dagger}_{\vec{\hat{h}}}  ) \ket{\mathrm{vac}},
\end{equation}
for which the subspace with one photon on either side of the PBS is entangled:
\begin{equation}
	\ket{\psi}_{\mathrm{anti-bunched}} \propto g_{1}g_{0}(   \ket{10} c^{\dagger}_{\vec{\hat{v}}}d^{\dagger}_{\vec{\hat{v}}} - \ket{01} c^{\dagger}_{\vec{\hat{h}}}d^{\dagger}_{\vec{\hat{h}}}  ) \ket{\mathrm{vac}}.
\end{equation}
However, when we measure the photons we have to ensure that neither we nor any other system has information about whether the photons are $\vec{\hat{h}}$ or $\vec{\hat{v}}$ polarized.  If this information is measured by any system then it will result in a projection into $ \ket{01}$ or $\ket{10}$.  For this reason we use another waveplate and PBS to measure modes $c$ and $d$ in the $45^{\circ}$ polarization basis $\{ (\frac{\vec{\hat{h}}+\vec{\hat{v}}}{\sqrt{2}}), (\frac{\vec{\hat{h}}-\vec{\hat{v}}}{\sqrt{2}}) \}$, which gives completely random outcomes for $\vec{\hat{h}}$ or $\vec{\hat{v}}$ photons.  Hence, a successful application of the Duan and Kimble protocol will project the matter system into a maximally entangled state\footnote{There may be a trivial phase depending on which detectors click.}.

Compared to single photon protocols, antibunching schemes are more sensitive to photon loss as they require two photons to successfully reach the detectors and may have a more complex apparatus and hence a higher loss rate.  However, photon loss only affects the success probability and not the fidelity.  Furthermore, since each term of the superposition contains a photon that traverses both arms of the apparatus, the scheme is insensitive to differences in path lengths.

\subsubsection{The double-heralding scheme}
\label{sec:double_herald}

\begin{figure}[t]
\centering
\includegraphics{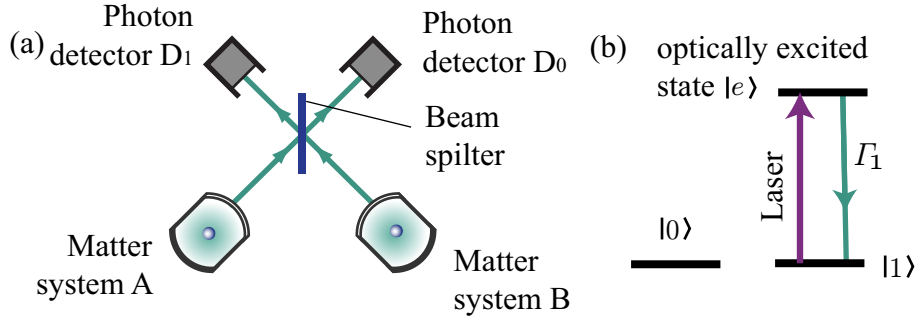}
 \caption{An outline of Barrett and Kok's protocol for generating entanglement by double heralding the parity measurement made by a single photon scheme. (a) the experimental apparatus is similar to that in the Cabrillo proposal, shown in Fig.~{\ref{fig:Cabrillo}}, but with the removal of the frequency filters. (b) an $L$-level electronic structure; $\ket{0}$ and $\ket{1}$ are degenerate, or nearly degenerate, in energy; The excited state $\ket{e}$ decays to $\ket{1}$ via a photon emission, whereas decay/absorption between $\ket{0} \leftrightarrow \ket{e}$ is not permitted due to a selection rule.}
\label{fig:Double-Herald_Outline}
\end{figure}

We now turn to the double-heralding scheme\cite{BK01a}, which uses matter qubits with an $L$-level structure arranged as shown in Fig.~{\ref{fig:Double-Herald_Outline}}.  The procedure consists of four steps: preparation, first heralding, bit flips and second heralding:
\begin{enumerate}
    \item \textit{Preparation}: Both qubits are initialized in the $\vert + \rangle = (\ket{0}+ \ket{1})/\sqrt{2}$ state;
    \item \textit{First heralding}: Both qubits are subjected to a $\pi$-pulse of frequency $\omega$, which causes the transition $\vert 1 \rangle \rightarrow \vert e \rangle$.  Next, the photon detectors $D_{1}$ and $D_{0}$ wait for detector clicks until enough time has passed that it is very unlikely that the cavity still contains any excitations.  We proceed to the next step of the protocol only if we have seen exactly one click.
    \item \textit{Bit flips}: Both qubits are bit flipped by the rotation $X_{A}X_{B}$;
    \item \textit{Second heralding}: We repeat the $2^{\mathrm{nd}}$ step. If and only if we again see one click then the protocol has succeeded.
\end{enumerate}

Notice that the first two steps are the same as for the Cabrillo protocol when $\theta= \pi/4$ and decay is only possible by one channel. Hence, after the first heralding we have a mixed state:
\begin{equation}
	 \left( \frac{2}{4-T} \right) \kb{\Psi^{\phi}}{\Psi^{\phi}} + \left( \frac{2-T}{4-T} \right) \kb{11}{11} .
\end{equation}
After the bit flip we have
\begin{equation}
	 \left( \frac{2}{4-T} \right) \kb{\Psi^{-\phi}}{\Psi^{-\phi}} + \left( \frac{2-T}{4-T} \right) \kb{00}{00}.
\end{equation}
When we perform the second heralding, the erroneous term can not emit any photons and so can not lead to a detector click.  Furthermore, on the second heralding the photon traverses the opposite path to that of its first journey, and hence both terms of the superposition acquire the same phase due to path length difference\footnote{This assumes that path length has not varied between the first and second round.}.  Although, in any one successful round only a single photon is detected, the whole protocol is successful after two photons have been detected.

Double heralding is not just suitable for producing Bell pairs, but can also be used to produce a parity projection for growing graph states (see section~{\ref{sec:Parity_Projection}}).  For graph state growth we omit the preparation step, and apply the protocol directly to the graph state qubits.  Upon success we undo the $X_{A}X_{B}$, and the result is an odd parity projection.
 
\subsection{A scaled up device}

\begin{figure}[t]
\centering
\includegraphics{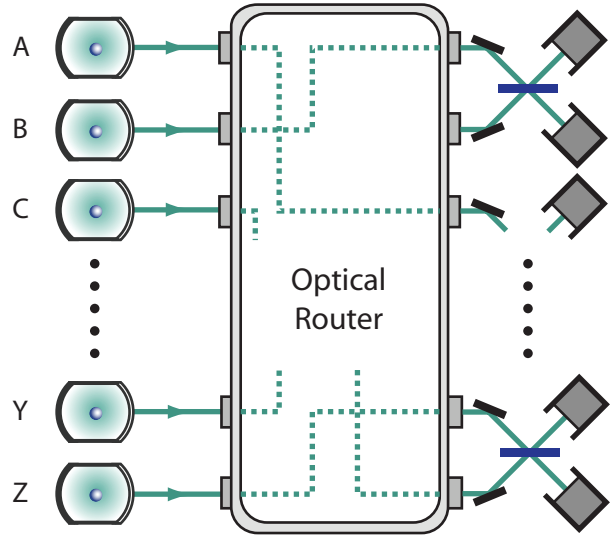}
\caption{Schematic of a DQC: Nodes of a DQC are alphabetically labelled, and each node contains some number of qubits of which at least one is optically active. Within a node control fields can manipulate qubits to high fidelity, so each node is miniature quantum computer.  However, large scale quantum computing requires entanglement generation between nodes.  Photon decay paths from different nodes are switched by an optical router, enabling entanglement generation between any two nodes.}
\label{fig:Router}
\end{figure}

To recap, in this section we have reviewed some protocols for generating entanglement and the physical mechanisms they use.  Although entanglement is essential for quantum computing, quantum states with bipartite entanglement alone are not sufficient.  Rather we require that many nodes of a distributed quantum computer can be entangled with each other.  Therefore, all of the entanglement proposals will require an optical multiplexer that switches the paths of photons, as illustrated in Fig.~\ref{fig:Router}.  

In the case of double heralding we are able to grow graph states by omitting the preparation step.  When successful, the double heralding procedure produces an entangling parity projection, but when it fails it projects two qubits into a separable states.  These success and failure outcomes have a simple graphical interpretation in terms of either connecting graph vertices or removing two vertices from the graph.  Even though double heralding grows graphs probabilistically, we see in section~{\ref{sec:Parity_Projection}} that strategies exist for efficient graph growth.

However, what holds for double heralding will not automatically transfer over to other schemes for generating entanglement.  For example,  $\Lambda$ level schemes require us to initialise in some separable state that can decay to either logical state, and this loss of prior entanglement during the preparation is unavoidable. Thus, to make use of such protocols it is necessary to have at least one additional qubit present at each node, where the two (or more) qubits within a node interact in such a way that we can perform good quality quantum gates between them. Then the  optically active qubits can negotiate entanglement, while the secondary qubits act as the logical qubits\cite{DB01a,DBMM01a,BBFM01a}. It is well known that once a Bell pair is shared between two nodes it can be used to implement an entangling two-qubit gate, and hence enable either circuit model quantum computing or one-way quantum computing.  When using ancillary qubits in this fashion the graph state is also immune from damage, reducing the need for strategies for probabilistic graph growth, but at the cost of additional node structure.

\section{Growth strategies}
\label{sec:Growth_strategies}

\subsection{Graph growth with parity projections}
\label{sec:Parity_Projection}

\begin{figure}[t]
\centering
    \includegraphics[width=\columnwidth]{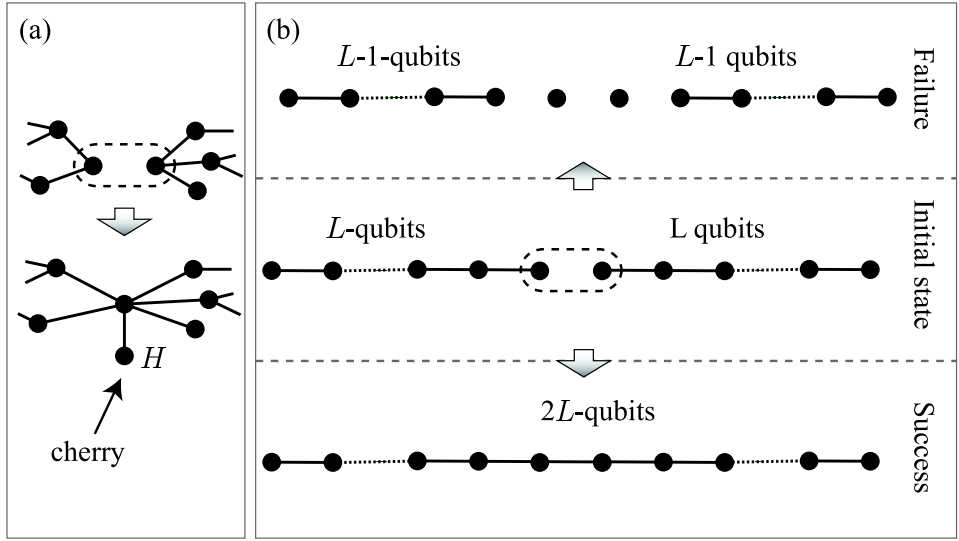}
    \caption{Examples of graph growth by parity projection: (\textit{a}) the outcome of a parity projection on two qubits both part of a graph state with many neighbouring vertices.  One qubit, say $A$, acquires all of the neighbours of the other qubit, $B$.  Whilst qubit $B$ becomes connected to qubit $A$ and only qubit $A$, with the addition of a Hadamard rotation. We can remove this Hadamard to obtain a constructively defined graph state.  When a qubit only has 1 neighbour, as qubit $B$ does, we call this qubit a \textit{cherry}.  Note that parity projections do not have a basis towards one qubit, we could equally describe the resulting state with qubit $A$ as the cherry.  Since the roles of $A$ and $B$ are interchangeable they are said to be redundantly encoded. (\textit{b}) the growth a \textit{chain} of qubits with a probabilistic control-$Z$ operation.  Provided an \textit{offline} resource of chains greater than some critical length, the longest available chain will, on average, grow in length.}
    \label{fig:graph_growth}
\end{figure}

From the constructive definition of a graph state, it trivially follows that if we can perform $CZ$ gates and local operations then we can construct graph states.  However, many proposals for entanglement generation are projective rather than unitary, especially when the entanglement is induced by measurements. Indeed, to our knowledge the Repeat-Until-Success protocol\cite{LBK01a} is the only measurement based protocol for generating entanglement that upon success generates a true control-$Z$ gate as opposed to a projection.  Whereas most of the entanglement protocols for distributed quantum computing only succeed when we detect some odd parity photon measurement signature, which we might expect corresponds to an odd parity projection on some qubits $x$ and $y$:
\begin{equation}
 	P^{-} = \kb{01}{01}_{x,y} + \kb{10}{10}_{x,y},
\end{equation}
where we use the negative sign to denote odd parity because it can also be written in term of Pauli-operators:
\begin{equation} 
 P^{\pm} = \left( \id \pm Z_{x} Z_{y} \right)/2,
\end{equation}
which also defines the even parity projector, $P^{+}$.  Since $\pm Z_{x}Z_{y}$ is a stabilizing operator, this notation shows that $P^{\pm}$ project onto the subspace stabilized by the $\pm Z_{x}Z_{y}$ of the same sign.  Since such parity projectors project onto a stabilizer state we might expect that they are suitable for constructing graph states.  Indeed, in Fig.~{\ref{fig:graph_growth}a} we illustrate how a parity projection affects two qubits $A$ and $B$: they become \textit{redundantly encoded}, which is LU-equivalent to a graph state with qubit $A$ gaining all of the neighbours of qubit $B$, whilst $B$ becomes a dangling \textit{cherry} with $A$ as its only neighbour.  

To verify that this is the effect of an odd parity projection, we first consider the initial state of the constructive defined graph state, $\ket{\mathscr{G}_{i}}$, which we divide into three partitions:  qubit $A$, qubit $B$, and the remaining graph state $\ket{\mathscr{G}'}$: 
\begin{equation}
 \ket{\mathscr{G}_{i}} = \frac{1}{2}(\ket{0}+\ket{1}\mathscr{Z}_{A})_{A}(\ket{0}+\ket{1}\mathscr{Z}_{B})_{B}\ket{\mathscr{G}'}, 
\end{equation}
where $\mathscr{Z}_{x}$ denotes a product of all the phase-flip operators resulting from the application of control-$Z$ gates involving graph vertex $x$, or formally:
\begin{equation}
 \mathscr{Z}_{x} = \prod_{y \in N(x)} Z_{y}.
\end{equation}

It follows that projection into the odd parity subspace generates:
\begin{eqnarray}
 \ket{\mathscr{G}_{f}} & \propto & P^{-} \ket{\mathscr{G}_{i}}, \\ \nonumber 
  & = &  \frac{1}{\sqrt{2}}(\ket{0, 1}\mathscr{Z}_{B}+\ket{1, 0}\mathscr{Z}_{A})_{A, B}  \ket{\mathscr{G}'}, \\ \nonumber
  & = &   \frac{1}{\sqrt{2}}( \mathscr{Z}_{B} X_{B} H_{B}  ) (\ket{0, +}+\ket{1, -}\mathscr{Z}_{A}\mathscr{Z}_{B})_{A, B}  \ket{\mathscr{G}'} ,\\ \nonumber
  & = &   \frac{1}{\sqrt{2}}  ( \mathscr{Z}_{B} X_{B} H_{B}  ) \left( CZ^{A}_{B} (\ket{0}+\ket{1}\mathscr{Z}_{A}\mathscr{Z}_{B})_{A} \ket{+}_{B} \right)  \ket{\mathscr{G}'}, 
\end{eqnarray}
where removal of the local unitaries $\mathscr{Z}_{B} X_{B} H_{B}$ clearly gives the constructively defined graph state depicted in Fig.~{\ref{fig:graph_growth}a}.  Note that we have only indicated the local unitary $H_{B}$ in the figure. Throughout this review we will omit Pauli-operators from our illustration of states LU equivalent to some graph state.  We use this convention because the presence of a Pauli-operator does not dramatically effect the evolution of a graph state under Pauli measurements, whereas the presence of a Hadamard rotation does have a significant effect\footnote{For example, a graph state qubit with a Hadamard measured in the $X$-basis will generate the same final state as a $Z$-basis measurement on an equivalent graph state without the Hadamard present.  In contrast Pauli-operators are only capable of more interchanging the outcomes of Clifford group measurements.}. Note that an even parity projection would give the same state \textit{up to} a difference in local Pauli operators.

\subsection{Probabilistic graph growth}
\label{sec:Probabilistic_graph_growth}

A broad range of entanglement protocols, generating control-$Z$ gates or parity projections, enable graph growth.  However, entanglement generation in distributed quantum computers will typically be probabilistic, with a failed attempt projecting the qubits onto some separable state.  This raises the question of whether probabilistic graph state growth can be performed efficiently.  A naive approach to graph growth would restart whenever it fails to entangle any two qubits.  Denoting the probability of generating entanglement by $p$, the overall probability of  success for this naive strategy will fall exponentially as $p^{n}$, where $n$ is the number of required number of attempts at entanglement generation.  An exponential fall in success probability entails an exponential amount of time until success, and hence we lose any speed-up gained from using a quantum algorithm.  Therefore, we say that a strategy for probabilistic graph growth is efficient when the temporal and spatial overheads scale polynomially in the size of the graph state.

Indeed numerous efficient strategies do exist, and they all exploit the way failed entanglement generation causes only local damage to graph states\cite{N01a,DR01a,BK01a,B01a}.  Typically, the exposition of these strategies has assumed that entanglement is produced by a control-$Z$ gate, though it is trivial to modify them for graph growth by probabilistic parity projection.  For the entanglement generation protocols that we are interested in, failure to entangle two qubits will always either: project the qubits into a separable state of the computational basis \textit{or} produce an unknown phase error on the qubits.  Therefore, we can measure out these qubits in the $Z$-basis, applying a correction to their neighbours where necessary\footnote{A $\ket{0}$ result requires no correction, whereas a $\ket{1}$ outcome requires that that all graph neighbours are phase-flipped.}.  We say that failure \textit{damages} the two qubits, and that they have to be \textit{reset} before we attempt to regrow the lost graph vertices.  Note that if a failure could cause bit-flip noise, we would also have to reset their neighbours\footnote{More formally, if qubit $x$ is involved in a failed procedure resulting in bit-flip noise then we would have to reset all qubits $y$ such that $y \in N(x)$.  Duan and Raussendorf assumed this worse case scenario when evaluating their graph growth strategy\cite{DR01a}. Therefore, when we describe their strategy in the next section there will be slight differences in some expressions}.   This property of graph states is closely related to a trick introduced into linear optical quantum computing by Knill \textit{et al}\cite{KLM01a}, which teleports successful instances of probabilistic gates into the main algorithm\cite{YoranReznik}.

\subsubsection{Dynamic growth strategies}

\begin{figure}[t]
\centering
    \includegraphics[width=\columnwidth]{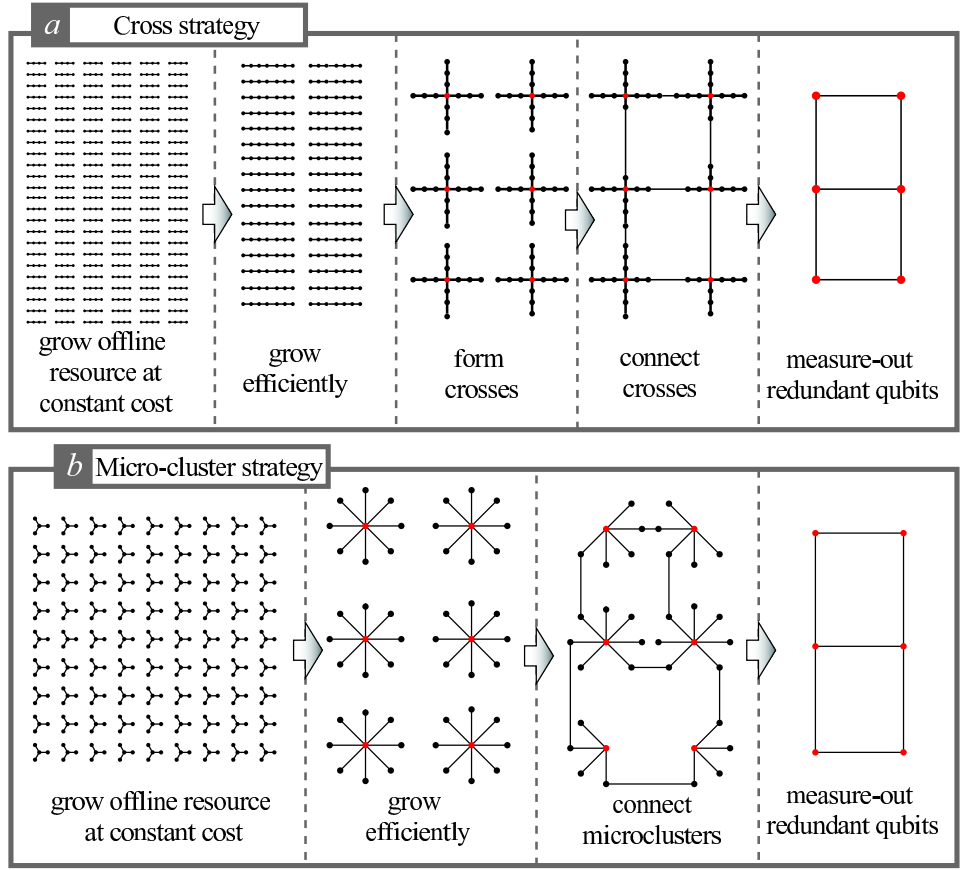}
 \caption{Examples of Re-routing growth strategies: (\textit{a}) the \textit{cross} strategy proposed by Duan and Raussendorf: (\textit{b}) the \textit{micro-cluster} strategy proposed by Nielsen.  For an idea of the number of qubits required, the number of qubits/connections lost at each stage of graph growth has for a success probability of $p=1/3$.}
    \label{fig:Piece_meal_growth_strategies}
\end{figure}

Here we outline two strategies that we call \textit{dynamic} graph state growth strategies,  so-called because they do not use a predecided sequence of attempts at generating entanglement.  Implementing these protocols would require our quantum computer to be able to attempt entangling operations between any, or almost any, pair of qubits.  First we consider the growth of long chains of qubits, which is the first stage of graph growth for the proposal of Duan and Raussendorf\cite{DR01a}.  In Fig.~{\ref{fig:graph_growth}b} we show the failure and outcome possibilities when connecting two chains, each initially containing $L$ qubits.  This succeeds with probability $p$, and the successfully joined chain has length $2L$.  However, if it fails, then the result is two chains of length $L-1$.  From this, we can calculate the expectation value of the longest available chain: $\langle L_{f} \rangle = p.2L + (1-p)(L-1)$.  For, on average, growth of the chain, the expected final chain length must be larger than the initial chain, $\langle L \rangle < \langle L_{f} \rangle$.  This simplifies to requiring that the initial chains are longer than some critical length $L_{c}= p^{-1}-1$.  Growing these critical length chains can be expensive for small $p$, but the resource cost does not scale with the size of the computation.  Therefore chain preparation is a constant cost that does not affect the efficiently of the strategy, and we call such states \textit{offline} resources.  Furthermore, Duan and Raussendorf showed the time to prepare chains of length $L$ scales as $\log_{2}(L-L_{c})$.  From here Duan and Raussendorf propose that chains are connected at the middle to produce a large cross shaped graph state, as outlined in Fig.~{\ref{fig:Piece_meal_growth_strategies}a}.  The central vertex of this cross will be present in the final target graph state, the purpose of all other qubits is simply to buffer against graph damage.  When attempting to connect two crosses the probability of an outright failure drops exponentially with the size of the buffer. After a successful connection, any remaining buffer qubits are removed by Clifford group measurements in the $X$ or $Y$ basis. By constructing suitable sized cross shaped graphs states, the chance of failing any connection in building an $N$ by $N$ square lattice, or cluster state, is kept to a small probability at subexponential cost in time and qubits\cite{DR01a}.  
A strategy in a similar spirit is the \textit{micro-cluster} strategy proposed by Nielsen\cite{N01a}, which we outline in Fig.~{\ref{fig:Piece_meal_growth_strategies}b}.  Instead of cross shaped graph states, this strategy proposes star shaped objects known as micro-clusters, of which only the central vertex will be present in the final graph state.  This strategy has a similar scaling to that of Duan and Raussendorf. Note that micro-cluster are LU-equivalent to GHZ states, named after Greenberger, Horne and Zeilinger who first described them\cite{GHZ01a}; and as such the terminology often used interchangeably.

\subsubsection{Percolation strategies}
\label{sec:percolation}

\begin{figure}[t]
\centering
    \includegraphics[width=\columnwidth]{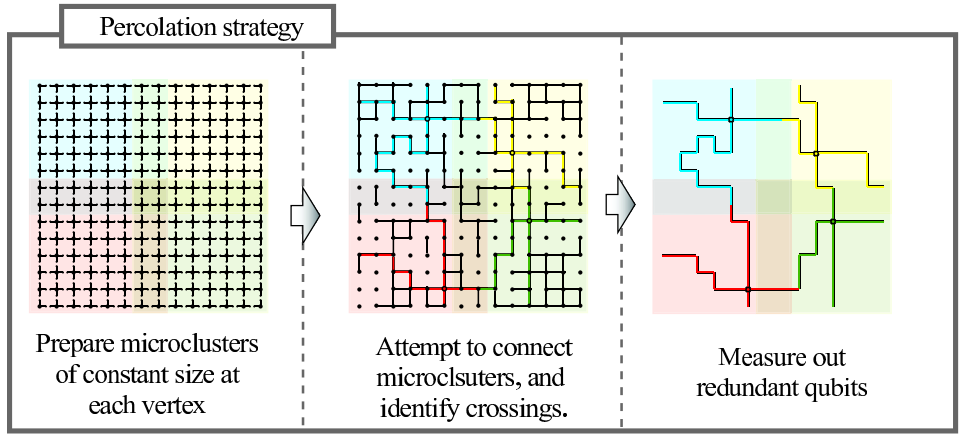}
    \caption{A simple example of graph state growth by percolation. First we produce microclusters arranged in a grid, and then attempt to join the microclusters into a square grid, much like the strategies in Fig.~{\ref{fig:Piece_meal_growth_strategies}b}.  However, rather than ensuring that the microclusters are large enough that every connection is almost certain to succeed, many will fail.  With the grid divided into blocks of qubits (we show 4 blocks with different colours which slightly overlap), we require that a qubit can be found in each block such that a path of successful connects leads from that qubit to neighbouring qubits.  When the successful probability is greater than some threshold (1/2 for square lattice and 5 qubit microclusters), it is asymptotically certain that this strategy succeeds.}
    \label{fig:percolation_strategies}
\end{figure}

Both the aforementioned strategies assume that the quantum computer is capable of attempting an entanglement operation between almost any two qubits, with the decision making process made dynamically by an assisting classical computer.  Building a quantum computer that provides the flexibility of being able to change which qubits we entangle is clearly more technologically challenging than one that does not.  A natural question is whether graph states can be grown when the quantum computer can only entangle each qubit with a fixed number of predecided qubits?  Indeed it has been shown that \textit{percolation} phenomena can be exploited to grow graph states when the interconnectivity of a device is limited\cite{KRE01a,ACL01a,BEFMMS01a}.  In Fig.~{\ref{fig:percolation_strategies}} we illustrate a simple percolation strategy where no qubit needs to be entangled with more than 4 other qubits. The overall square lattice is sub-divided into partially overlapping square blocks that will each represent one qubit in the final graph state.  Each block is successful if there is a uninterrupted path from left-to-right and top-to-bottom, and we also require crossings in the overlapping region.  Assuming that such paths exist, all qubits except one in each block are measured in a Pauli-basis to reduce down to a square lattice.  The \textit{percolation} phenomena that they exploit is well known in other fields and describes phase changes in problems such as the movement of solvents through filter paper (chromatography) and the movement of petroleum through fractured rock.   Kieling~\textit{et al}\cite{KRE01a}  utilise results from bond-percolation theory that to show that provided the probability of a hole defect is smaller than some critical value, then at a polynomial overhead the success probability is asymptotically certain\footnote{If the target graph state is a square lattice of $L$ by $L$ qubits, and we use blocks of order $L^{\mu}$ by $L^{\mu}$ qubits ( $\mu>0$), then the probability of success approaches 1 as $L \rightarrow \infty$.}.  For a square lattice this critical probability is $1/2$, though more favourable probabilities can be tolerated by utilising different lattice topologies, such as 3D square lattices and diamond lattices, and by simply increasing the size of the initial microclusters.  Percolation phenomena has also been investigated when sites rather than bonds fail\cite{BEFMMS01a}, and in the context of quantum communication\cite{ACL01a}.

\subsubsection{Problems with probabilistic graph growth}

All the above strategies take the probability of successful entanglement generation, $p$, as a constant, and assume perfect fidelity operations.  Given these constraints it is claimed that graph growth can be performed efficiently.  However, as $p$ decreases the overhead in time and qubits required increases faster than $1/p$.  Indeed, extremely low $p$ makes graph growth impossible when the expected time to make a single successful connection exceeds the decoherence time\footnote{This is a measure of the rate at which noise takes a qubit into a randomized state.} of the qubits.   Paying heed to the effects of noise, as we increase the number of qubits used in a single cross, micro-cluster or percolation block, we increase the amount of noise that will eventually collect on a single graph state qubit\cite{RRM01a}.  Hence, under realistic conditions probabilistic graph growth strategies will be too noisy for quantum computing when $p$ is too low.  Exact bounds are not known, but what is clear is that significant gains can be reaped by generating entanglement with a lower probability of damaging the underlying graph. 

\section{Robust computation}

While the growth strategies discussed in the previous section allow us to deal with probabilistic entangling operations, more effort is necessary to deal with noise accumulated during construction and measurement of graph states. Such errors may be caused by errors in the local operations or entangling operation, for example due to photon detector dark counts, or simply by decoherence of the matter qubits used to store the graph state. In any case, any scalable architecture must take steps to suppress such noise in order to maintain a finite probability of obtaining the correct output from a computation.

\subsection{Error correction and fault-tolerance}

One of the major successes of quantum information theory has been the development of a formal theory of fault-tolerant computation. In such a model, the quantum information is encoded using an error correction code to protect against interaction with the environment and other random errors. A universal set of gates is then constructed in such a way that errors are not amplified. A regular regime of error correction between logic gates ensures that the computation remains in the encoded subspace. Provided that the error rate is below a critical value, known as the fault-tolerance threshold, errors can be suppressed indefinitely with only polynomial resource overhead. The exact value of the fault-tolerance threshold depends on several factors including the specific error correction code used, the structure of the fault-tolerant logic gates and on the underlying physical operations available.
 
Maximising the fault-tolerance threshold is an extremely important open problem as it allows for computation to be performed in noisier systems, reducing the technological overhead required to implement quantum computation. Although many families of quantum error correcting codes have been proposed, determining a threshold is far from trivial. A number of different factors affect the threshold for fault-tolerance. Important factors include how many errors can be detected and diagnosed by an error-correcting code and how many gates are required per round of error-correction.

While standard error correction codes, such as the Shor and Steane codes\cite{Sho95a,Ste96a}, can be applied to one-way quantum computing, it is far from clear whether codes designed for the circuit model of quantum computation will prove to be the best option in this new architecture\cite{Nielsen05a,raussendorf-2005-,aliferis-2006-73}.

One promising proposal, brought forward by Raussendorf\cite{raussendorf-2005-}, is to use surface codes\cite{Kitaev03a} to protect against errors. Surface codes represent a topological approach to error correction, in which a tiling of local operators on some surface are used to stabilize the subspace in which the computation takes place. An advantage of this approach is that it finds a natural substrate in three dimensional extension of cluster states. Such topological approaches to fault tolerance provide a fertile area for further research and may well result in higher fault tolerance thresholds. 

\subsection{Entanglement distillation}

We have briefly reviewed how fault tolerant quantum computation enables reliable computations despite the threat of errors occurring at any stage, e.g. graph state growth, storage, or measurement.  However, these techniques can only cope with small error rates.   But what if noise is significantly worse at a particular stage, such as during our entangling operation used for graph state growth.  This possibility naturally occurs in a distributed architecture, where entangling operations between nodes will occur via a fundamentally different mechanism than inside nodes.   For hybrid matter/optical systems the entangling operation may be noisy dues to photon loss, dark counts\footnote{Where photon detectors click in absence of a photon.}, imperfect alignment of optical elements or non-identical photon sources.  Whilst there will also be noise processes within nodes, it seems that these must strictly be less noisy than the appropriate fault tolerance threshold.  Provided that noise is low inside nodes, many uses of a noisy inter-node entangling operation can be used to produce a single high fidelity entangling operation. The family of protocols that achieve this are aptly called \textit{entanglement distillation} protocols, and can provide high fidelity graph state growth even when the entangling operation is significantly nosier than any fault-tolerance threshold.

Such a powerful tool does not come for free and requires some additional complexity in the structure of quantum computer.  To employ entanglement distillation protocols requires more than one qubit per node, with the precise number of qubits depending on the type of noise and protocol employed.  Most of the literature focuses on so-called Pauli noise, where the entangling operations produces a Bell state up to random Pauli Z (phase), X (bit) or Y errors.    D\"{u}r and Briegel\cite{DB01a} showed that 4-5 qubits per node can be used to distill almost any entangling operation that suffers from Pauli noise.  Later Jiang \textit{et al}\cite{JTSL02a} proposed a different scheme that distilled general Pauli noise using 4 qubits per node, but only needed 3 qubits per node for phase noise.  Both these protocols can be employed both within the one-way model and the circuit model,  as they use a one qubit buffer per node to store a high fidelity Bell pair which is then consumed to perform any high fidelity logical operation.  Exploiting the simplicity of graph state growth, it has been shown\cite{C01a,CB01a} that the need for a one qubit buffer can be eliminated, with  high fidelity graph states growth possible using only 2 (phase noise or photon loss noise) or 3 (general Pauli noise) qubits per node.  A description of the internal working of these protocols is beyond our current scope, but we hope to have conveyed that a great deal of inter-node noise can be distilled with a node containing a modest number of qubits.  Note that,  if much larger nodes are available then we may use more complex protocols for graph state distillation\cite{ADB05a,GCR01a}. 

\section{Conclusions}
In this introduction we have seen how hybrid matter-optical systems can be combined with the one-way model of computation to yield an extremely powerful and robust architecture for scalable quantum computing. As such technologies mature, additional local structure will allow for higher efficiency, by buffering growth and distilling the entanglement present. These systems hold the promise of universal control without the limitations imposed by local architectures, and may offer a shortcut on the route to scalable devices. 

\section{Acknowledgements}
This review contains portions of the doctoral theses of Earl T. Campbell and Joseph Fitzsimons, and so we thank the following people who made useful comments on the relevant chapters: Pieter Kok, Brendon Lovett,  Vlatko Vedral, Erik Gauger, Thomas Close and Yuichiro Matsuzaki.  Earl T. Campbell is currently supported by the Royal Commission for the Exhibition of 1851, and was supported by the QIP IRC during his doctorate. Joseph Fitzsimons is supported by Merton College.

\end{document}